\documentclass[12pt,final]{iopart}

\usepackage[english]{babel}
\usepackage[T1]{fontenc}
\usepackage{times}
\usepackage{amssymb}
\usepackage{verbatim}
\usepackage[dvipdfmx]{graphicx}
\usepackage{mediabb}
\usepackage{overpic}
\usepackage{booktabs}
\usepackage{float}
\usepackage[usenames,dvipsnames]{color}

\usepackage{hyperref}
\hypersetup{
  colorlinks,
  citecolor=Blue,
  linkcolor=Blue,
  urlcolor=Blue,
  breaklinks=true
}

\newcommand{\bra}[1]{\langle #1\mid}
\newcommand{\ket}[1]{\mid #1 \rangle}
\newcommand{\OSMPS}{OSMPS}
\newcommand{\hc}{h_{\mathrm{crit}}}
\newcommand{\QI}{quantum Ising}
\newcommand{\LRQI}{long-range quantum Ising}
\newcommand{\CLRQI}{Long-Range Quantum Ising}
\newcommand{\hatH}{H}
\newcommand{\hatsig}{\sigma}
\newcommand{\eqref}[1]{(\ref{#1})}

\newcommand{\csm}{Department of Physics, Colorado School of Mines, Golden,
  Colorado 80401, USA}
\newcommand{\jila}{JILA, NIST and University of Colorado, Boulder, 
  Colorado 80309-0440, USA}
\newcommand{\jdw}{Department of Physics \& Astronomy, Rice University,
  Houston, Texas 77251, USA}
\newcommand{\jhu}{The Johns Hopkins University Applied Physics Laboratory,
  Laurel, MD, 20723, USA}

\begin{document}

\title{Critical Phenomena and Kibble-Zurek Scaling in the Long-Range Quantum
  Ising Chain}

\author{Daniel Jaschke$^1$, Kenji Maeda$^1$, Joseph D.\ Whalen$^{1,2}$,
  Michael L.\ Wall$^{1,3}$\footnote{Present address: \jhu},
  and Lincoln D.\ Carr$^1$}
\address{$^1$ \csm}
\address{$^2$ \jdw}
\address{$^3$ \jila{}}


%

\begin{abstract}

We investigate an extension of the quantum Ising model in one spatial dimension
including long-range $1 / r^{\alpha}$ interactions in its statics and dynamics
with possible applications from heteronuclear polar molecules in optical
lattices to trapped ions described by two-state spin systems. We introduce
the statics of the system via both numerical techniques with finite size and
infinite size matrix product states and a theoretical approaches using a
truncated Jordan-Wigner transformation for
the ferromagnetic and antiferromagnetic case and show that finite size
effects have a crucial role shifting the quantum critical point of the external
field by fifteen percent between thirty-two and around five-hundred spins. We
numerically study the Kibble-Zurek hypothesis in the long-range quantum Ising
model with Matrix Product States. A linear quench of the external field
through the quantum critical point yields a power-law scaling of the defect density as
a function of the total quench time. For example, the increase of the defect
density is slower for longer-range models and the critical exponent changes by
twenty-five percent. Our study emphasizes the importance of such long-range
interactions in statics and dynamics that could point to similar phenomena in
a different setup of dynamical systems or for other models.

\end{abstract}

\submitto{\NJP}
\maketitle

\section{Introduction                                                           \label{sec:intro}}

Common quantum many-body models, such as the \QI{} model or Hubbard models,
include only a nearest-neighbor interaction, while e.g. the Coulomb forces or
the dipole-dipole
interactions have power law tails of the form $1 / r^{\alpha}$. This leads to
the core question: "to what extent does the physics of these models changes when
including decaying long-range interactions?" Physical models with tunable
long-range interactions have been successfully realized in recent experiments due
to advances in atomic and molecular physics. This especially includes
experiments in heteronuclear polar molecules in optical lattices \cite{Carr2011,%
DeMillePT2015} and trapped ions \cite{Blatt2012}. Further examples for the
realization of a long-range Ising model are the cold ion experiment in
\cite{Bohnet2016} realizing a two dimensional lattice and ultracold molecules
with long-range dipole interactions, which have been successfully realized in
recent experiments \cite{Neyenhuis2012,Molony2014}. Furthermore, realizations
in Rydberg atoms through long-range van der Waals interactions \cite{Zeiher2015,
Labuhn2016} or solid state physics, e.g. based on $\mathrm{LiHoF}_4$
\cite{Silevitch2010}, exist and provide a large number of spins, e.g. in
comparison to ultracold ion experiments at the current stage. In addition,
magnetic dipoles can lead to the same long-range interactions which have been
realized in recent experiments in terms of Chromium \cite{Naylor2016}, Erbium
\cite{Chomaz2016}, and Dysprosium \cite{Tang2015}, and further setups with
long-range interactions tuned by optical cavities have been made
\cite{Dogra2016}, too. The list of experiments can be further extended
with nitrogen vacancies in diamonds with a ferromagnetic long-range Ising
interaction \cite{Wei2015}. Recent advances in quantum annealing have fostered
the mapping of problems onto a generalized quantum Ising model
\cite{Lucas2014}, which is a more general Hamiltonian than the long-range
interaction. This is as well the case for quantum spin glasses \cite{Wu1991}.
These systems have in common
that they can be described by two-state spin models with $1/r^{\alpha}$
interactions coupled to external transverse fields, including the long-range
quantum Ising (LRQI) model.
While the experimental systems range from implementations in one to
three dimensions, we study a one-dimensional system, because our numerical
techniques are best suited to this case.
The recent work of several groups discussed new
emerging phases due to long-range interactions \cite{Landig2016} and different
regimes that hold or violate the Lieb-Robinson bounds, limiting the speed of transferring
information depending on the initial
state \cite{Santos2016}.

For numerical studies of many-body systems, tensor network methods offer
a variety of algorithms which can be tailored to certain problems
\cite{Schollwoeck2011}. The infinite Matrix Product States (iMPS)
\cite{McCulloch2008} on infinite translationally invariant lattices and the
Matrix Product States (MPS) for finite systems
are well suited for the ground state calculations of quantum systems. Dynamics
can be simulated with the
Time-Dependent Variational Principle (TDVP)
\cite{Haegeman2016}, where other methods as Krylov or local Runge-Kutta
were explored for comparison \cite{WallNJP2012,Zaletel}. The \QI{} model
is not only fundamental because of the various experimental implementations
mentioned, but also since it has a similar classical model in the Ising model,
originally introduced in 1925 \cite{Ising1925} with analytic solutions in
one and two dimensions. Moreover, the quantum model in $d$
dimensions can be mapped onto the classical Ising model in $d + 1$ dimensions;
the solution for the classical model in two dimensions is linked to our
one dimensional quantum chain.
Therefore, it is an ideal starting ground to explore long-range
interactions before including them into other models and
has always been the toy model to explore a huge variety
of methods before using them on other models since it is the description for a
vast majority of physical problems. This includes methods such as the
exact spectrum, perturbation theory, or the mapping of the ferromagnetic
case to the corresponding classical Ising model in higher dimensions without
using numerical methods. It is applicable to many problems that can be mapped into
two-level systems and therefore accessible for researchers from fields as different as
solid state physics and atomic, molecular, optical physics just mentioned before.

While the experimental systems can be
in higher dimension, we consider one dimensional spin chains
with long-range interactions subjected to the external magnetic field in its
transverse direction. From there, we lay out the discussions of static and dynamic
simulations via tensor network methods for models with long-range interactions.
Numerical algorithms based on tensor networks perform optimally in one dimension
and have limitations in higher dimension, although they have been proposed for
two dimensional systems \cite{Verstraete2008}.
However, the Kibble-Zurek
mechanism yields a fascinating topic for the \LRQI{} model in real time
evolution; for example, quenching across the critical point could be used for
state preparation whenever it is easier to prepare the ground state in one
phase.

\begin{figure}[htbp]
  \centering
  \includegraphics[width=3.2in]{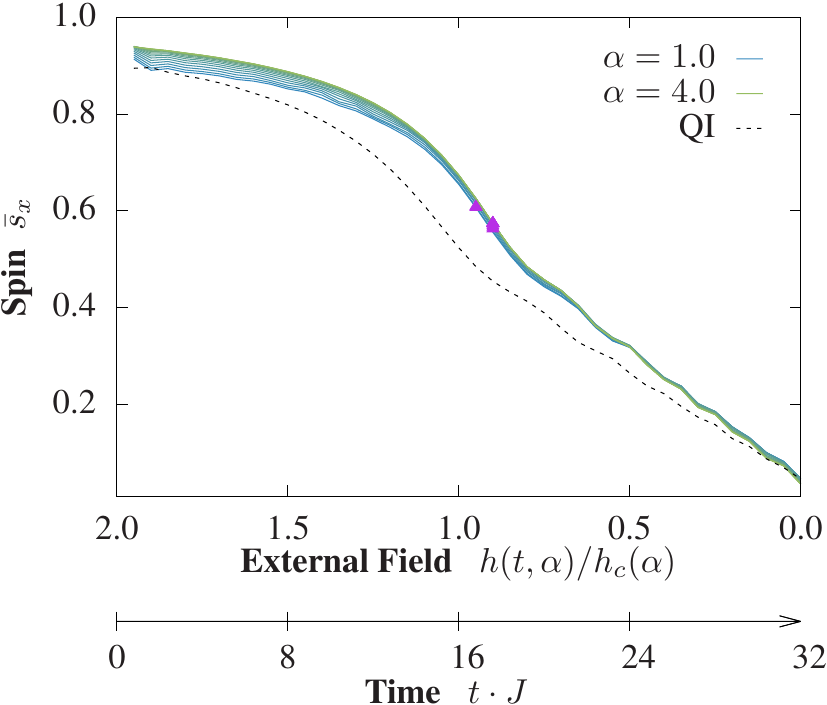}
  \caption{\emph{Quench of the \CLRQI{} Model.}
    The mean local $x$-magnetization as average over the sites of the
    measurement of $\sigma_{i}^{x}$, i.e., $(1 / L) \sum_{1}^{L} \sigma_{i}^{x}$,
    is a typical observable for spin
    models and realizable in cold ion experiments.
    The measurement during the quench from the
    paramagnetic phase to the to the ferromagnetic phase for different
    interaction strength governed by power law $ 1 / |j - i|^{\alpha}$ displayed
    for $\alpha$ in steps
    of $0.2$ is compared to the nearest-neighbor \QI{} (QI, dashed curve).
    The triangles mark spots with first maximum of the gradient disregarding
    numerical fluctuations. The magnetization in $x$-direction is vanishing
    equally for all $\alpha$, since the quench is starting at
    $h(t=32, \alpha) = 2 \hc(\alpha)$ and ending at $h(t=\tau, \alpha) = 0$.
    $\hc$ is the value of the quantum critical point which differs for each
    $\alpha$. We emphasize the unified dynamics of the quench independent
    of $\alpha$ for the symmetric setup around the critical point.
                                                                                \label{fig:DynMeasFerro3d}}
\end{figure}

As for the usual quantum Ising (QI) chain, that is integrable, both static
properties such as the phase diagram, critical exponents or correlation lengths,
and dynamic properties have been studied at length. The latter includes the
Kibble-Zurek mechanism (KZM) being applied to show that the density of
defects in the ferromagnetic chain follows a scaling law for linear
quenches \cite{ZurekPRL2005,Cincio2007,delCampo2012}, i.e. the density of
defects increases for faster
quenches across the quantum critical point. Those properties of the Ising model
have always paved the road to study other models. Then, there appears a natural
question: how do these long-range interactions affect the well-known results
in the \QI{} model? These answers are relevant to ultracold molecules with
characteristic long-range dipole-dipole interactions or Rydberg atoms with
long-range interactions.
Therefore, we present the reader new results on the Kibble-Zurek scaling with the
comparison of the static results using three different approaches. Our
analytical approach uses an approximation which truncates phase terms of the Jordan-Wigner
transformation and the two numerical algorithms cover infinite and finite
systems. With those comparisons we see finite size effects as the lower
critical external fields for decreasing system sizes;
dynamic simulations are based on these results. We introduce the dynamics
with experimentally measurable observables such as the local magnetization in
Fig.~\ref{fig:DynMeasFerro3d} and compare the scaling of defects with the
Kibble-Zurek hypothesis. The Kibble-Zurek mechanism posits that for decreasing
quench times, the defect density increases less for the ferromagnetic case in
regard to the nearest-neighbor \QI{} model. In contrast, the number of defects
stays at the same rate for the antiferromagnetic model for faster quenches
in comparison to the \QI{} model. With the exponent of the Kibble-Zurek
mechanism changing about $25\%$                                                 
in the ferromagnetic case, this should be an observable effect. Apart from
the prediction of defects when quenching across the quantum critical point
for preparing states it might even allow for characterization of long-range
effects according to the critical exponent.

Throughout the paper we concentrate on
both the antiferromagnetic and ferromagnetic case.
Moreover, we address the challenges with those studies, e.g. simulations
breaking down in certain regimes of the \LRQI{}. We use the
Open Source Matrix Product States package (\OSMPS{}) \cite{WallNJP2012} freely available from \cite{openMPS}.
As a secondary goal, we provide with those practical
calculations and discussions on the numerical results a basis for future
researchers to interpret their results on statics and dynamics gained with
\OSMPS{}
or any other tensor network method library. The models suitable for further
studies range from more advanced spin models of $XYZ$-type to all kind of
Hubbard models.

While static aspects such as the phase diagram of the \LRQI{} model have been studied
in the past \cite{Dutta2001}, the antiferromagnetic case has been
investigated recently by Koffel~\cite{Koffel2012} and extended in
\cite{Vodola2015} to the correlations. We add to these works the aspects of
infinite size MPS algorithms. A study in the thermodynamic limit with Linked
Cluster Expansions for both the ferromagnetic and antiferromagnetic case
has been presented in \cite{Fey2016}. The dynamics of the \LRQI{} chain has lately
attracted a huge interest. The entanglement growth of a bipartition during a
quench was described in \cite{Schachenmayer2013}. The speed of information
spreading and the building of correlations was studied in \cite{Eisert2013,%
Richerme2014,Hauke2013,Santos2016,Maghrebi2016} and the equilibration time
of long-range quantum spin models in \cite{Kastner2011}. We not only extend
the picture of dynamics in the
long-range interaction in the Ising model to the Kibble-Zurek hypothesis, but
provide a detailed path how such aspects can be studied numerically keeping a
side-by-side view of the antiferromagnetic and ferromagnetic cases.

The structure of the paper is as follows: after introducing the Hamiltonian
of the \LRQI{} model in Sec.~\ref{sec:ham}, Sec.~\ref{sec:theo_approx} presents the
theoretical approximation for the phase boundary.  The actual results are then
discussed in Sec.~\ref{sec:phaseboundary} containing the phase boundaries for
all different methods used, the observables during the dynamics, and the
Kibble-Zurek scaling for the \LRQI{} model. We finish with the discussion of
our results in the conclusion. The description of the numerical methods follows
in the \ref{sec:methods} and consists of the numerical algorithms used and the
studies on their convergence. Further aspects on the convergence can be found
in \ref{sec:convmps}.

\section{Modeling Long-Range Interactions in the Quantum Ising Model            \label{sec:ham}}

The system is modeled by the \LRQI{} Hamiltonian,
\begin{eqnarray}                                                                \label{eq:LRQI_H}
  \hatH_{\rm LRQI} &=& 
  - J \sum_{i<j} \frac{\hatsig^z_i\hatsig^z_j }{ |i-j|^{\alpha} }
  - |J| h \sum_i \hatsig^x_i
  ~,
\end{eqnarray}
where $i$, $j$ are 1D lattice coordinates, $J$ and the unitless $h$ represent the strengths
of the interaction and transverse field, respectively, and $\alpha$ takes some
positive value parameterizing the long-range interactions. Examples for
well-known models are Coulombic interactions with $\alpha = 1$,
dipole-dipole interactions decreasing with $\alpha = 3$, and $\alpha = 6$
for van der Waals models and the tail of the Lennard-Jones potential.
In the following, we
investigate phase diagrams both for the ferromagnetic ($J>0$) and
antiferromagnetic ($J<0$) cases in Eq.~\eqref{eq:LRQI_H}. The \LRQI{} model in
Eq.~\eqref{eq:LRQI_H} reduces to the usual \QI{} Hamiltonian in the limit of
$\alpha\to\infty$,
\begin{eqnarray}                                                                \label{eq:QI_H}
  \hatH_{\rm QI} &=&
  - J \sum_{i} \hatsig^z_i \hatsig^z_{i+1}
  - |J| h \sum_i \hatsig^x_i
  ~, 
\end{eqnarray}
which exhibits a quantum phase transition at $h = 1$ both in ferro- and
antiferromagnetic cases. For $h > 1$ we have in both cases the paramagnetic
phase where spins align with the external field in the limit $h \to \infty$.
The ferromagnetic phase appears for $h < 1$ where the
ground state is described for $h = 0$ by a $\mathbb{Z}_{2}$ symmetric ground
state $\left(\ket{\uparrow \ldots \uparrow} \pm \ket{\downarrow
\ldots \downarrow} \right) / \sqrt{2}$ subject to spontaneous symmetry
breaking.
The antiferromagnetic ground state has a staggered order in $z$-direction
$\left(\ket{\uparrow \downarrow \uparrow \downarrow \ldots} \pm
\ket{\downarrow \uparrow \downarrow \uparrow \ldots} \right) / \sqrt{2}$
referred to as N\'eel order. The Hamiltonian of the \LRQI{} in
Eq.~\eqref{eq:LRQI_H} builds the foundation of the following work and the
standard \QI{} from Eq.~\eqref{eq:QI_H} is used frequently to check limits.

A common approach to quantum many-body systems are critical exponents. These
universal scaling laws are independent of the system size and describe quantities
as a function of the distance to the quantum critical point \cite{SachdevQPT}.
One typical example for such a quantity is the correlation length in the system.
We relate to the critical exponents in the analysis of the Kibble-Zurek scaling,
where \cite{Polkovnikov2011} gives a review of the link between the critical
exponents and the crossing of a quantum critical point in a quench. This setup
is described by the critical exponent $\nu$ and the dynamical
critical exponent $z$. The more detailed description is included in
\ref{sec:dynamics}.

\section{Theoretical Boundary via a Truncated Jordan-Wigner Transformation      \label{sec:theo_approx}}

Having a general overview of the \LRQI{} model, we now derive a
well-controlled analytic approximation to rely on. This is important to have
a comparison to the numerical results introduced later.
In general, the \LRQI{} problem is not exactly solvable like the
nearest-neighbor case, but a truncation of the phase terms in the Jordan-Wigner
transformation regains a long-range fermion model along the Kitaev model, which
allows us to approximate the critical transverse field for quantum phase
transitions, $\hc$.
We show the derivation of the critical transverse fields with the use of the
Jordan-Wigner transformation \cite{Jordan1928,SachdevQPT} and a truncation
approximation. The derivation can be divided into two steps. First, we use
the Jordan-Wigner transformation to map our spin Hamiltonian onto fermions.
The Jordan-Wigner transformation is a non-local mapping from spins onto
fermions, and fermions onto spins, where the non-locality of the transformation
is necessary to obey the corresponding commutation relations. In the case of a
nearest-neighbor spin Hamiltonian, the result of the Jordan-Wigner transformation
is a nearest-neighbor fermion Hamiltonian where the non-local terms of the
transformation cancel each other. In contrast, long-range spin Hamiltonians
are mapped into long-range fermion Hamiltonians with $k$-body interactions of
arbitrarily large $k$.
After the truncation in the second step, we obtain a Kitaev-type Hamiltonian
with long-range two-body interactions, which has been solved exactly in
\cite{Vodola2014}. The Kitaev Hamiltonian is itself important as it describes
spinless fermions with p-wave pairing interactions and displays topological
states \cite{Kitaev2001}.
For a complete picture we show both steps to derive the
theoretical boundary. In the following, we rewrite the \LRQI{} Hamiltonian
in one dimension from Eq.~\eqref{eq:LRQI_H} with a general site dependent
coupling $V(i, j)$:
\begin{eqnarray}                                                                \label{H_gQI}
  \hatH &=& \sum_{i<j}^{} V(i,j) \, \hatsig_i^z \hatsig_j^z
  - |J| h \sum_{i}^{} \hatsig_i^x ~,
\end{eqnarray}
and keep a constant in front of the external field in order to have a unitless
$h$. This general equation describes a variety of different models starting
from a quantum spin glass with random $V(i, j)$ originally proposed as
classical model \cite{Sherington1975}, over the completely connected case with
$V(i, j) = 1$ for all $i, j$ which is equivalent to a special case of the
Lipkin-Meshkov-Glick model \cite{Lipkin1965,Dutta2010} ranging to \emph{chimera}
setup in the d-wave quantum annealing computers \cite{Boothby2016} with a
limited number of interaction to other qubits.
This Hamiltonian reduces to our \LRQI{} model with $V(i,j)=-J/|i-j|^{\alpha}$,
while to the usual \QI{} model with $V(i,j) = -J \delta_{j-i,1}$. In order to
use the Jordan-Wigner transformation, we have to rotate our coordinate system
around the $y$-axis mapping $\sigma^{x} \to \sigma^{z}$ and $\sigma^{z} \to
- \sigma^{x}$ following the approach used in \cite{SachdevQPT}. We obtain
\begin{eqnarray}                                                                \label{eq:H_gQIrot}
  \hatH_{\mathrm{rot}} &=& \sum_{i<j}^{} V(i,j) \, \hatsig_i^x \hatsig_i^x
  - |J| h \sum_{i}^{} \hatsig_i^z ~,
\end{eqnarray}
where the minus sign of the transformation cancels out appearing as a pair
in the interaction term. The Jordan-Wigner transformation, which is exact
for the usual \QI{} model, is given by
\begin{eqnarray}
\hatsig_i^{x}&=&
\biggl\{\prod_{j<i} (1-2c_j^{\dagger}c_j) \biggl\} (c_i+c_i^{\dagger})
~,\\
\hatsig_i^{y}&=&
-{\rm i} \biggl\{\prod_{j<i} (1-2c_j^{\dagger}c_j) \biggl\} (c_i-c_i^{\dagger})
~, \\
\hatsig_i^z&=&
1-2c_i^{\dagger}c_i
~, 
\end{eqnarray}
where $c_i$, $c_i^{\dagger}$ are annihilation and creation operators for
spinless fermions satisfying the anticommutation relations. 
%
%
Applying the Jordan-Wigner transformation, we can rewrite the Hamiltonian
Eq.~(\ref{eq:H_gQIrot}) with the fermionic operators,
\begin{eqnarray}
\hatH&=&
\sum_{i<j}^{}V(i,j)(c_i^{\dagger}-c_i)
\biggl\{\prod_{m = i+1}^{j-1} (1-2c_m^{\dagger}c_m)\biggl\}
(c_j+c_j^{\dagger})
\nonumber \\
&&
- |J| h\sum_{i}^{} (1-2c_i^{\dagger}c_i)
~.
\end{eqnarray}
Note that for the nearest-neighbor potentials, like in the usual \QI{} model,
there are no operator products in terms of the index $m$, thus the Hamiltonian
becomes quadratic in the fermionic operators, reducing to the well-known
Jordan-Wigner transformed quantum Ising model. The approximation step replaces
the operator products $\prod_{m} (1 - 2 c_i^{\dagger} c_i)$ in the interaction
term of the Hamiltonian running over the index $m$ with an identity. This is
equivalent with truncating fermionic operators up to quadratic order to obtain
an analytically solvable form and the reason behind calling it a truncated
Jordan-Wigner approach. We continue with
\begin{eqnarray}                                                                            \label{H_trun}
\hatH&\approx&
\sum_{i<j}^{} V(i,j)
(c_i^{\dagger}c_j+c_j^{\dagger}c_i+c_i^{\dagger}c_j^{\dagger}+c_jc_i)
\nonumber \\
&&
- |J| h \sum_{i}^{}  (1-2c_i^{\dagger}c_i)
~.
\end{eqnarray}
This Hamiltonian corresponds to the long-range Kitaev model exactly solved
in \cite{Vodola2014}. Our truncation approximation becomes exact only for
the nearest-neighbor potentials. In general, it becomes better for
ferromagnetic states, since they satisfy
$\langle \hatsig_i^z \rangle =  \langle 1 - 2c_i^{\dagger} c_i \rangle
\approx 1$, where $\hatsig_i^{z}$ refers to the rotated Hamiltonian in
Eq.~\eqref{eq:H_gQIrot}.






We perform a standard Fourier transform and Bogoliubov transform \cite{SachdevQPT},
which completes the diagonalization of Eq.~\eqref{H_trun} with the result,
\begin{eqnarray}
\hatH&=&
\sum_{q} \epsilon_q
\left(\gamma_q^{\dagger}\gamma_q-\frac{1}{2}\right)
~,
\end{eqnarray}
where the excitation spectrum $\epsilon_q$ is given by 
\begin{eqnarray}                                                                            \label{exspec}
  \epsilon_q
  &=& 2 \sqrt{\left({\cal S}_q \right)^2 + \left({\cal C}_q \right)^2
      + 2h {\cal C}_q + |J| h^2}~, 
\end{eqnarray}
with cosine and sine transforms of the potential where we replaced $(i, j)$
with the distance $r = j - i$ valid for the long-range quantum Ising model
\begin{eqnarray}
{\cal C}_q&=& 
\sum_{r=1}^{\infty} V(r)\cos(qr)
~,\\
{\cal S}_q&=&
 \sum_{r=1}^{\infty} V(r)\sin(qr)
~, 
\end{eqnarray}
respectively. The excitation spectrum Eq.~\eqref{exspec} could have a gapless
$\Gamma$-point ($q=0$) at the critical field
\begin{eqnarray}
\hc^{\Gamma}
&=&
- \frac{1}{|J|}
\sum_{r=1}^{\infty}
V(r)
~,
\end{eqnarray}
and a gapless $K$-point ($q=\pm\pi$) at
\begin{eqnarray}
\hc^{K} &=& - \frac{1}{|J|} \sum_{r=1}^{\infty} (-1)^{r} \, V(r)~,
\end{eqnarray}
though it depends on details of the potentials as to whether or not the
spectrum exhibits the gapless points. For ferromagnetic long-range interaction,
$V(r)=-J/r^{\alpha}$ with $\alpha>1$, we have a gapless $\Gamma$-point at 
\begin{eqnarray}                                                                \label{eq:hc_FM}
  \hc^{\Gamma} &=& \sum_{r=1}^{\infty} \, \frac{1}{r^{\alpha}}
  ~=~ \zeta(\alpha) ~,
\end{eqnarray}
where $\zeta(\alpha)$ is the Riemann Zeta function. In contrast, for
antiferromagnetic long-range interaction, $V(r)=|J|/r^{\alpha}$, we have a
gapless $K$-point at
\begin{eqnarray}                                                                            \label{eq:hc_AFM}
\hc^{K}
&=&
\sum_{r=1}^{\infty}
\frac{(-1)^{r+1}}{r^{\alpha}}
~=~
\biggl(1-\frac{1}{2^{\alpha-1}}\biggl)\zeta(\alpha)
~.~~
\end{eqnarray}
We remark that for the usual \QI{} model with $V(i,j) =
- J \delta_{j-i,1}$, we expect a gapless $\Gamma$-point at $\hc^{\Gamma} = 1$
in the ferromagnetic case, and a gapless $K$-point at $\hc^{\Gamma} = 1$
in the antiferromagnetic case.

\section{Phase Diagram of the Long-Range Quantum Ising Chain                    \label{sec:phaseboundary}}

\begin{figure}[htbp]
  \centering
  \includegraphics[width=0.9 \columnwidth]{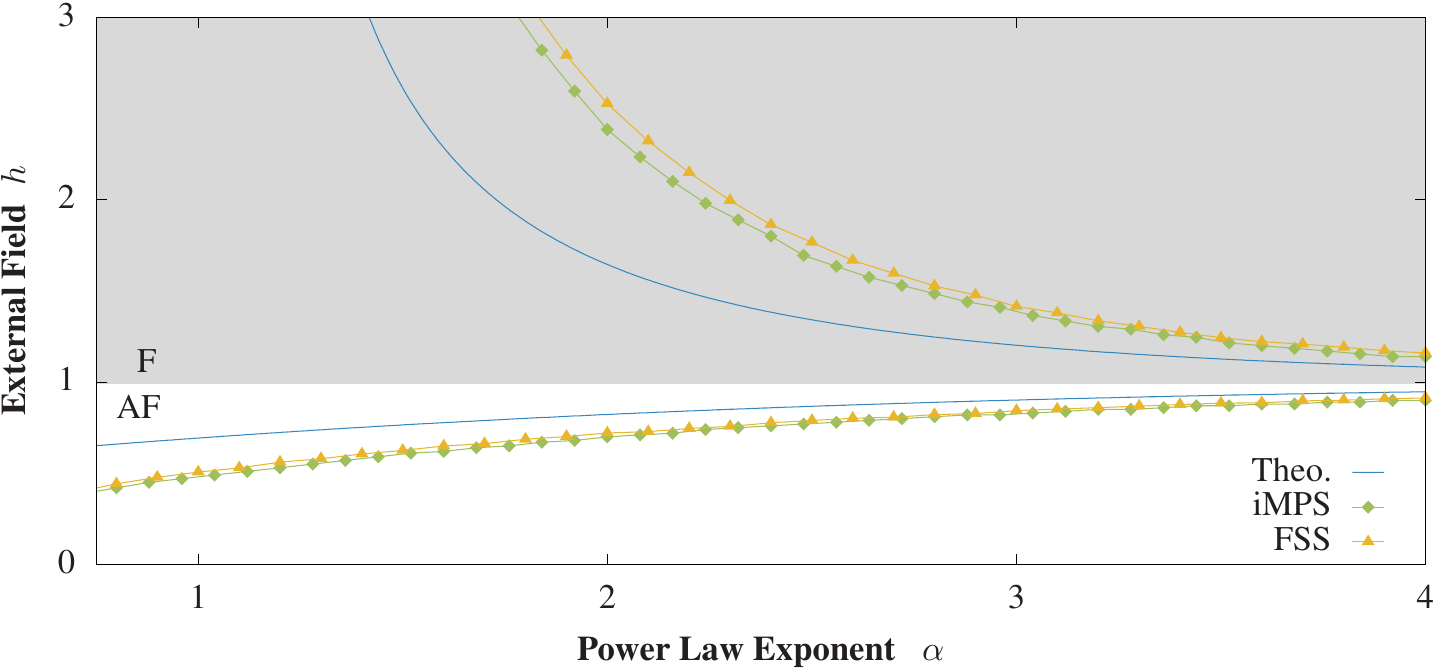}
  \caption{\emph{Phase Boundary.} Critical behavior of the long-range quantum
    Ising model in the ferromagnetic and antiferromagnetic case as a function
    of the power law exponent $\alpha$. We compare the theoretical
    approximation (Theo.) based on a truncated Jordan Wigner transformation,
    the infinite MPS results (iMPS), and the finite size scaling (FSS) of
    the MPS results. The boundary between the gray and white shading is the
    critical point in the limit $\alpha \to \infty$ corresponding to the
    nearest-neighbor quantum Ising model. We observe that both cases
    react differently when increasing the long-range interactions. In the
    ferromagnetic case the paramagnetic phase decreases at cost of the
    ferromagnet, while the antiferromagnetic phase becomes smaller in the
    other case due to the concurrency in the staggered order. All three
    different methods show these trends, although they show in the regime
    $\alpha$ significant relative differences above $0.1$.
                                                                                \label{fig:PhaseBoundaries}}
\end{figure}

Our first result for the phase boundary is obtained analytically.
In Sec.~\ref{sec:theo_approx} we used the Jordan-Wigner transformation, which
solves the nearest-neighbor \QI{} chain exactly, for the \LRQI{} leading to an infinite series
of fermionic operators. We truncate such series to obtain analytic estimates for
the critical magnetic fields and recall the results from Eqs.~\eqref{eq:hc_FM}
and \eqref{eq:hc_AFM}
\begin{eqnarray}
  \hc^{\rm FM}(\alpha) = \zeta(\alpha)~, \qquad
  \hc^{\rm AFM}(\alpha) = 
  \left(1 - \frac{1}{2^{\alpha-1}} \right)\zeta(\alpha)~,                       
\end{eqnarray}
for the ferromagnetic and antiferromagnetic cases, respectively. We compare
this to the numerical values of the critical field $\hc$ discussed
thoroughly in \ref{sec:statics}. In Fig.~\ref{fig:PhaseBoundaries} we
show the phase boundaries for the different methods, where the results of
the Kitaev model from Sec.~\ref{sec:theo_approx}
corresponds to the solid blue curves. We observe as a first trend that in the
ferromagnetic \LRQI{} chain, as $\alpha$ decreases, that is, as the potential
becomes more strongly long-range, the ferromagnetic phase becomes more favorable: the
negative energy contribution from the ferromagnetic ordering increases in
its absolute value. On the other hand,
in the antiferromagnetic \LRQI{} chain, the antiferromagnetic phase becomes
less favorable for smaller values of $\alpha$ because the long-range potential
induces more frustration in the staggered ordering. Next, we perform numerical
simulations to estimate the critical lines more accurately.

During the numerical simulations we follow two different paths. First we use
the infinite Matrix Product State algorithm
\cite{McCulloch2008,WallNJP2012} to search for a translationally invariant
ground state for a unit cell of $q$ sites. We search for this kind of ground
state for $q = 2$, although the Hamiltonian in Eq.~\eqref{eq:LRQI_H} fulfills
translational invariance in the thermodynamic limit for any $q$.
The energy is minimized via
variational methods and the strong finite size effects at the boundaries due to the long-range
interactions are avoided. Second, we consider finite size systems with a number
of sites $L \in \{ 32, 64, 128, 256, 512 \}$ and approximate the critical
point in the thermodynamic limit $L \to \infty$ via finite size scaling. On the one hand this is an appropriate
check on the results. On the other hand, we use finite size systems for
the dynamics later on and provide a prediction for finite size effects commonly
present in quantum simulators.
To identify the critical point we keep $\alpha$ and $J$ fixed and iterate over
the transverse field $h$. The von Neumann entropy as a function of the external
field has a maximum located at the critical value of the external field $\hc$.
For the nearest-neighbor model the critical point can be found based that the
area law for entanglement is violated for gapless states \cite{Eisert2010},
where the entanglement is measured with the von Neumann entropy. Although
long-range models may in general violate this area law away from the critical point,
we take the same approach knowing that the ground state obeys the area law for
the two limiting cases $h \to 0$ and $h \to \infty$.
The Ising model has a closing gap around the critical point and therefore the
von Neumann entropy $S$ can be used as indicator for the quantum critical point
\begin{eqnarray}                                                                \label{eq:vonNeumann}
  S(h) = - \sum_{i=1}^{\chi} \lambda_i \log \left(\lambda_i \right) \, ,
\end{eqnarray}
where $\lambda_i$ are the singular values squared of the Schmidt decomposition
of the system and $\chi$ is the maximum bond dimension controlling the
truncation of the Hilbert space. In the finite size systems, the eigenvalues of
the reduced density matrices correspond to $\lambda_{i}$ when splitting into a
bipartition. The entropy and the entanglement have a unique extremum at the
phase transition which must coincident with the closing gap and can therefore
be used to identify the critical point. An alternative approach would be to
use the correlation length of the system or the maximal gradient in the
magnetization, both with respect to the external field.
We explore the critical behavior of the infinite chain (triangles) and the
finite size scaling results (diamonds) in Fig.~\ref{fig:PhaseBoundaries}.

In conclusion, the relative difference between the iMPS and the theoretical
estimate is bound by $0.37$ ($0.17$) for the ferromagnetic (antiferromagnetic)
model considering $\alpha \ge 2$. The relative difference for two arbitrary
values $x$ and $x'$ is defined as $(x - x') / ((x + x') / 2)$. The
actual trend with the difference between iMPS and the theory curve vanishing
in the limit $\alpha \to \infty$ reflects the analysis in
Sec.~\ref{sec:theo_approx}. Further we state that the relative difference
between iMPS and finite size scaling results are below $0.11$ ($0.04$). The
details on the finite size scaling are described in \ref{sec:statics}.

\section{Scaling of Defect Density for the Kibble-Zurek Hypothesis under a
  Quantum Quench                                                                \label{sec:dynamics}}

\begin{figure}[htbp]
  \centering
  \includegraphics[width=0.7 \columnwidth]{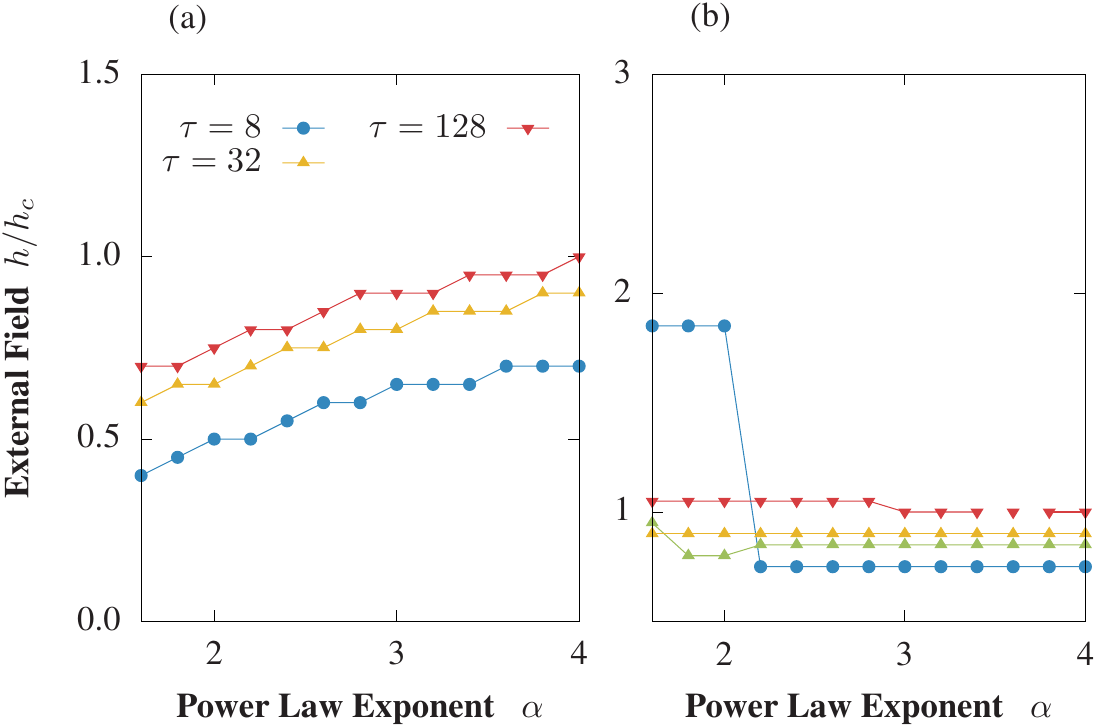}
  \caption{\emph{Dynamics and Phase Boundaries.} Around the critical value
    of the transverse field $\hc$, the gradient of the magnetization in the
    $x$-direction has its maximum. The system needs time to react to the
    change and therefore fast quenches with a low quench time $\tau$ reach
    the maximum in the gradient at later times. (a) The first maximum during
    the quench for the antiferromagnetic model. Instead of the time we show
    the corresponding transverse field $h(t)$ divided through the
    corresponding critical field. The small plateaus arise from the
    discretized measurements. (Remark: Small maxima (0.5 times max
    gradient) in the gradient are neglected.)
    (b) Analogous to (a) for the ferromagnetic case (same legend as (a)).
    The analysis failed for the data point with $\tau = 128$ and
    $\alpha = 3.6$, which was filled with the unconnected point from the
    simulation with $\chi = 95$. (Jump for fast quenches and small
    $\alpha$ appears for both bond dimensions).
                                                                                \label{fig:PhaseBoundariesDyn}}
\end{figure}

We now turn our
focus to the dynamics of the long-range quantum Ising model. We analyze the
quench through the quantum critical point by means of finite systems. We
first analyze measurable observables during the quench and then determine
the Kibble-Zurek scaling according to the final state, that is a power-law
describing the generation of defects as a function of the quench time and
a critical exponent. The linear quench for a time dependent external field
is defined as
\begin{eqnarray}
  h(t) = h_i + (h_f - h_i) \frac{t}{\tau} \, ,                                  \label{eq:linquenchh}
\end{eqnarray}
and replaces the coupling $h$ in Eq.~\eqref{eq:LRQI_H}. $\tau$ is
the total time of the quench. We quench from $h_i = h(t=0) = 2 \hc$ to
$h_f = h(t=\tau) = 0$. When considering our observables during the quench,
we stay close to experimentally realizable measurements. In
Fig.~\ref{fig:DynMeasFerro3d} we presented the average magnetization
$\bar{s}_{x}$
\begin{eqnarray}                                                                \label{eq:meansz}
  \bar{s}_{x} = \frac{1}{L} \sum_{k=1}^{L} \langle \sigma_{k}^{x} \rangle \, .
\end{eqnarray}
This is e.g. measurable in cold ion experiments \cite{Bohnet2016}.
We choose to plot the magnetization in the $x$-direction which is measurable
in the ferromagnetic and antiferromagnetic case, and can not average out in the simulations
due to superpositions, e.g. the ferromagnetic ground state
$\alpha \to \infty$, and $h=0$: $(| \uparrow \ldots \uparrow \rangle \pm |
\downarrow \ldots \downarrow \rangle ) / \sqrt{2}$ has no net magnetization
for a local measurement of $\sigma^{z}$.
We note in Fig.~\ref{fig:DynMeasFerro3d} for $\alpha \approx 4$ the steep
gradient around $t / \tau \approx 0.5$, which corresponds to the critical value
for the transverse field
in the ferromagnetic case treated here. Therefore, we extract the first maximum
of the gradient disregarding all minor fluctuations and plot them over $\alpha$
in Fig.~\ref{fig:PhaseBoundariesDyn}. We see a similarity to the critical value of
the transverse field from the static calculations for big $\tau$, while for
faster quenches the deviation from the static result increases. We present the
analogous analysis for the antiferromagnetic case as well in
Fig.~\ref{fig:PhaseBoundariesDyn}. For faster quenches the maximal gradient is
again below the static result leading to the assumption that this effect is due
to a delay during which the system takes time to react. The antiferromagnetic
case shows an additional effect as a function of $\alpha$: longer-range
interactions exhibit an extra delay in their maximum gradient. If we consider
the antiferromagnetic quench for $\tau = 128$ in
Fig.~\ref{fig:PhaseBoundariesDyn} (a), the maximum of the gradient moves from
$h = \hc$ for $\alpha = 4$ to values of $h \le 0.75 \hc$ for $\alpha < 2$. One
suggestion to explain this is the concurrent spin configuration in the
staggered order in the antiferromagnet which grows for smaller $\alpha$ and
might cause this extra delay to establish an antiferromagnetic state.
The results on this analysis vary if the quench scheme is not symmetric around
the critical point $\hc$.

\begin{figure}[htbp]
  \centering
  \begin{overpic}[width=0.65 \columnwidth,unit=1mm]{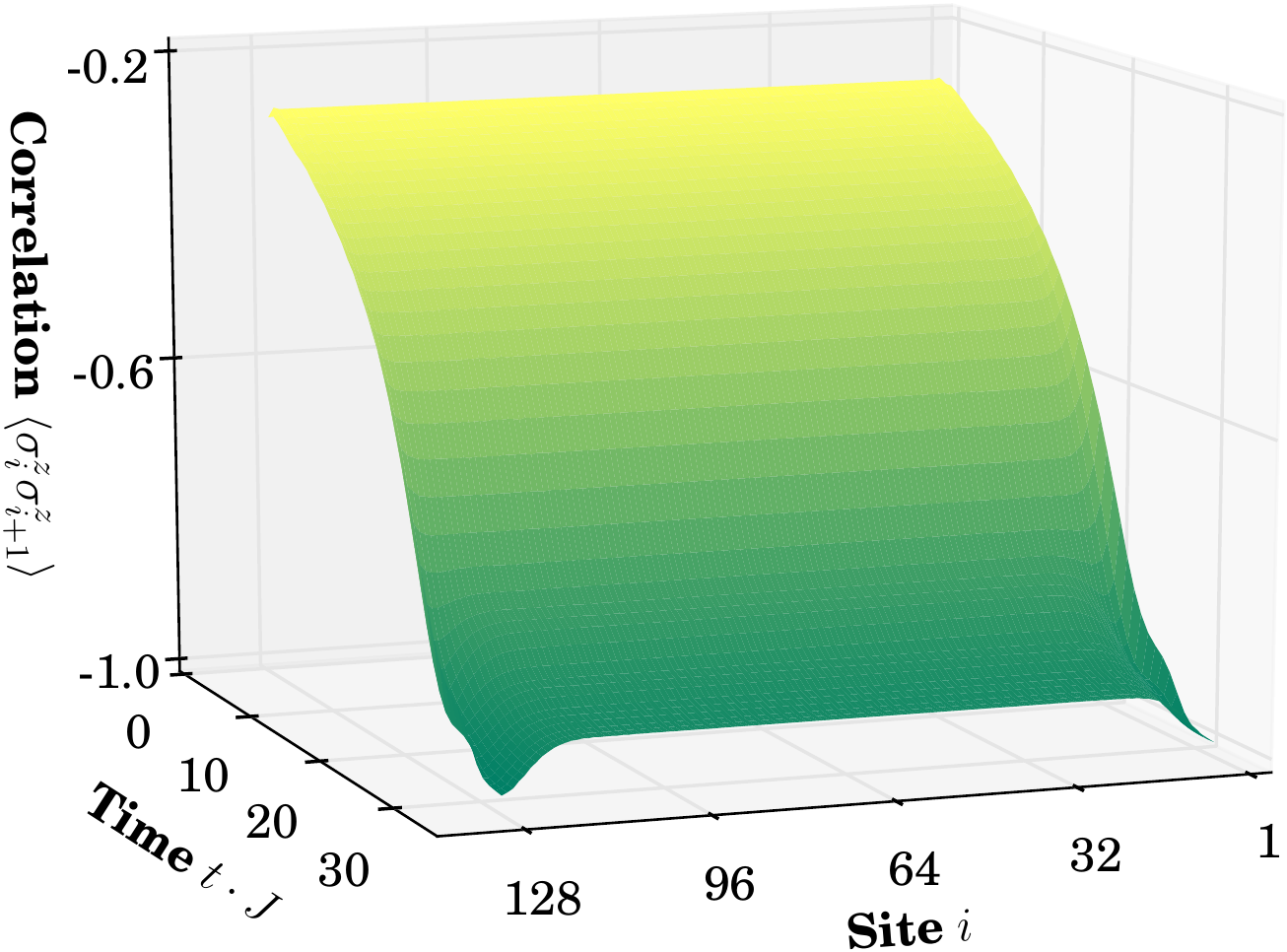}
    \put(39,9){{\color{blue}$\uparrow$}}
    \put(91,14){{\color{blue}$\uparrow$}}
  \end{overpic}
  \caption{\emph{Nearest-Neighbor Correlation in Dynamics.} Space-time evolution
    of the $z$ spin-spin correlation during the quench into the antiferromagnetic
    phase where the correlation between sites $i$ and $i + 1$ builds up as
    the external field vanishes.
    As expected for staggered order in an antiferromagnetic state the
    correlation is negative. Further we point out more strongly negative correlation due
    to boundary effects at the end of the quench, where the affected region is
    indicated with arrows ({\color{blue}$\uparrow$}).
                                                                                \label{fig:DynMeasAntiferroCorrxx3d}}
\end{figure}

Next, we present results for a non-global spin measurement. Such a measurement 
of a single spin in an optical lattice is technically possible \cite{Bloch2012}
and can be realized in cold ion experiments, too.
In Fig.~\ref{fig:DynMeasAntiferroCorrxx3d} we show the nearest-neighbor spin-spin
correlation in the $z$-direction for the antiferromagnetic case as a function of
time and site in the system. As mentioned above, local
measurements of $\hatsig^{z}$ average out due to symmetry apart from numerical
fluctuations. Correlations grow during the quench, and finite
system boundary effects are visible at both ends of the chain. The total
quench time is $\tau =32$, the system size $L=128$, and $\alpha=3$ in this
example, corresponding to dipole-dipole interactions.

\begin{figure}[htbp]
  \centering
  \includegraphics[width=0.9 \columnwidth]{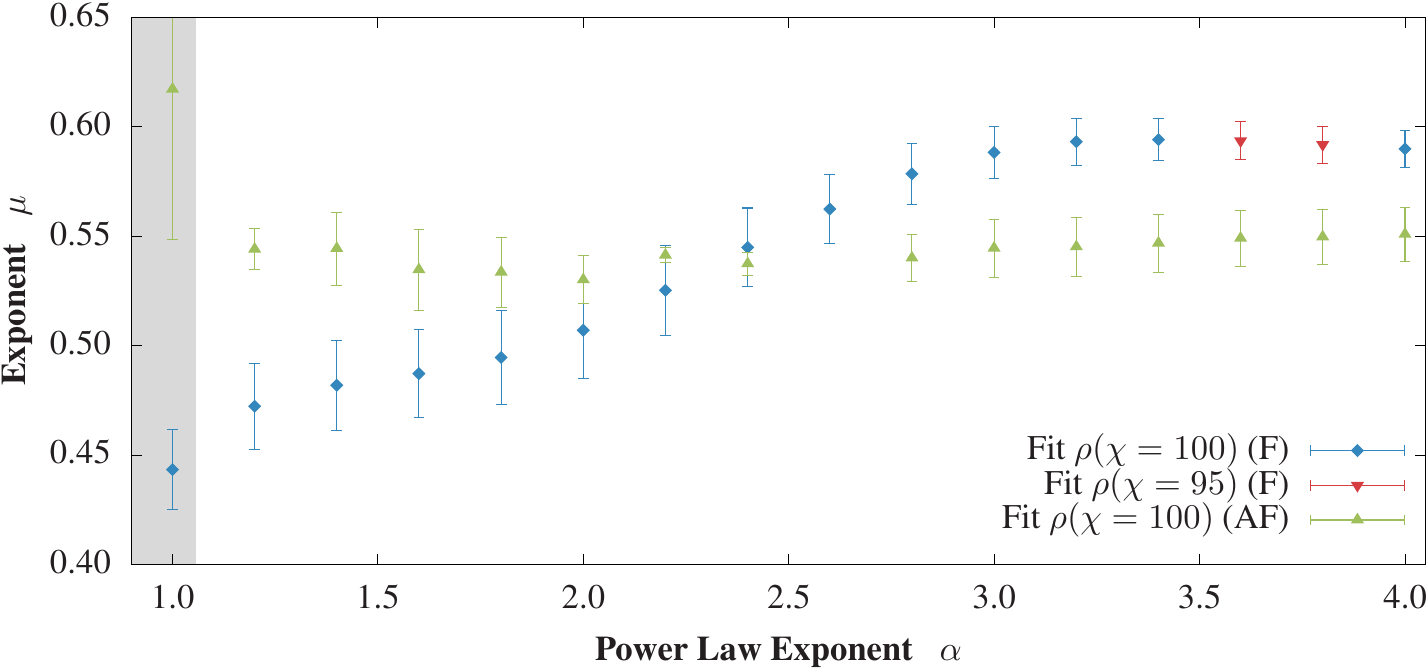}
  \caption{\emph{Kibble-Zurek Scaling for the \CLRQI{} Model.} The exponent
    $\mu$ relates the quench time to the defect density $\rho$ for the power law
    exponent $\alpha$ as in Eq.~\eqref{eq:kzmscaling}. The ferromagnetic model
    has smaller $\mu$'s for longer-range interactions. The defect density grows
    slower for longer-range interactions for faster quenches regardless of their
    original density. Two outliers in the data with large errors bars
    ($\alpha = 3.6$ and  $3.8$) have failed to converge and are replaced with
    simulations with lower bond dimension showing that there is no singularity.
    In contrast, for the
    antiferromagnetic model $\mu$ stays constant within the error bars
    and the defects increase at the same rate independent of long-range
    interactions, but do not necessarily have the same number of defects at
    the beginning. One non-converging point is not displayed ($\alpha = 2.6$).
    The error bars for both cases show the standard deviation from the fit
    of the critical exponent for the different quench times and do not contain
    truncation errors of the time evolution.
                                                                                \label{fig:KZMFerroEx}}
\end{figure}

Now we turn to the Kibble-Zurek hypothesis in the quenches already
discussed above. The Kibble-Zurek mechanism \cite{ZurekPRL2005,Dziarmaga2010,%
Polkovnikov2011} predicts a scaling of the defects in the quantum system as a
function of the quench time $\tau$. We recall that the scaling for the
nearest-neighbor Ising model is \cite{PolkovnikovCarr}:
\begin{eqnarray}                                                                \label{eq:nnkzmscaling}
  \rho \propto | \tau |^{- \frac{\nu}{z \nu + 1}} \, ,
\end{eqnarray}
where we use $\rho$ for the defect density, and $\nu$ and $z$ for the critical
exponents. We have already considered
a one-dimensional system in Eq.~\eqref{eq:nnkzmscaling}. Furthermore,
Eq.~\eqref{eq:nnkzmscaling} and the following analysis applies only to linear
quenches defined over $\partial h / \partial t = \mathrm{const.}$, but
extensions for the Kibble-Zurek scaling to non-linear quenches are possible
\cite{deGrandi2010}. Plugging in
the critical exponents for the nearest-neighbor quantum Ising model with
$\nu = 1$ and $z = 1$ we obtain $\rho \propto \tau^{- 0.5}$. The exponent $1/2$
is a checkpoint for the calculations in the nearest-neighbor limit
$\alpha \to \infty$. The match between theory and experiment does depend on
the experimental realization, but yields all over a power law for the defects
according to the review in \cite{Polkovnikov2011}.

We are interested in how much the exponent changes when tuning $\alpha$ for
the long-range interactions and how it deviates from the $1/2$
in the nearest-neighbor case. We estimate the exponent $\mu$ in the proceeding
analysis which is defined as
\begin{eqnarray}                                                                \label{eq:kzmscaling}
  \rho \propto \tau^{-\mu} \, .
\end{eqnarray}
Although the scaling can be used in more general cases, we restrict ourselves
to the linear quenches crossing the quantum phase transition located at the
critical value $\hc$ analyzed in the previous Section~\ref{sec:phaseboundary}.
This quench scheme starting at twice the critical value was discussed in
Eq.~\eqref{eq:linquenchh}. The quench starts in the paramagnetic phase and
ends in the ferromagnetic/antiferromagnetic phase.
When we are considering the ferromagnetic ground state, the number of defects
corresponds to the number of kinks, respectively domain walls,
and we use the defect density $\rho$
defined as the number of kinks per unit length and introduced in detail in
\ref{sec:methods}, Eq.~\eqref{eq:defectdensity}. Expressing $\rho$ in
terms of the kinks is constrained to the final point of the quench
$h = 0 = h_f$, where the measure is best defined, as the defects are clearly
identifiable in terms of the kinks.
The data in Fig.~\ref{fig:KZMFerroEx}, is obtained from simulations based on
systems with $L=128$ and evolution with the time-dependent variational
principle \cite{Haegeman2016} at $\chi = 95, 100$. We expect
that the critical exponent $\mu$ approaches $0.5$ in the limits
$\alpha \to \infty$ and $L \to \infty$ resulting in the ferromagnetic
nearest-neighbor quantum Ising model in the thermodynamic limit. We provide
a brief overview of the finite size effects in nearest-neighbor model in
\ref{sec:fssdyn} to support this statement. The error
bars are the standard deviation for $\mu$ of the fit for the different
values of $\tau$. Examples for the fit are shown in Fig.~\ref{fig:KZMFit} in
\ref{sec:dynamics}.
It does not contain the truncation or methodical error of the time
evolution method.
From the two curves with the different bond dimensions we deduce that certain
simulations fail, resulting in a large standard deviation in the fitting
procedure. We consider those points with large standard deviation as a failure
of convergence since another set of simulations with a slightly different bond
dimension fit into the trend of converged points and have an equally small
standard deviation.
The critical exponent grows for increasing values of $\alpha$, meaning
if we double the quench velocity for two systems with different long-range
interactions, the number of defects increases more slowly in the system
governed by longer-range interactions. But as we show in
\ref{sec:dynamics}, longer-range systems
have initially a higher defect density as shown in Fig.~\ref{fig:KZMFit}.

However, these results just discussed for the ferromagnetic case are not
true for the antiferromagnetic model. The critical exponent stays
constant independent of the long-range interaction. The deviation from the
nearest-neighbor case $\mu = 0.5$ should be due to finite size effects.
But we know from Fig.~\ref{fig:KZMFit} that the density of defects is again
higher for longer range interactions. The defect density of the
antiferromagnetic model is defined analogously to the ferromagnetic model: a
missing kink in the staggered order is one defect.

\section{Conclusion and Open Questions                                          \label{sec:conclusion}}

In this study we have established that the introduction of long-range
interactions changes the physics of one-dimensional quantum spin systems
non-trivially as observed in various aspects of the statics and the
defect density and therefore merits further attention. This is not limited to
the \QI{} model, e.g. the Bose-Hubbard and many other models could be
discussed under the influence of the long-range interactions such as
dipole-dipole interactions. First we summarize the results from our study of
the \LRQI{} model and then raise open problems associated with long-range
quantum physics.

We have presented multiple aspects of \LRQI{} model, e.g. the Kibble-Zurek scaling,
which lead to a better understanding of physical systems with long-range
interactions, e.g. the scaling of generated defects during the quench through
a quantum critical point. For the ferromagnetic case we find a slower
increasing defect density in comparison to the nearest-neighbor limit when
increasing the quench velocity. The defect density is higher in comparison to the
nearest-neighbor model for the same quench time. Our simulations give a detailed
picture of the critical value of the transverse field in the thermodynamic
limit using infinite Matrix Product States along with a practical guidance how to
resolve issues introduced by degeneracy in those systems. We also presented an
alternative to the infinite size algorithm using finite size scaling methods
to obtain the value of the critical external field in the thermodynamic limit.
This uncovers the finite-size effects, which will certainly be present in
experiments with small system sizes. The relative difference for the value
of the critical point between $L=32$ and $L=512$ is approximately $15\%$ for
the antiferromagnetic case for $\alpha = 3$. The same comparison for the
ferromagnetic yields a relative difference of around $20 \%$.
The iMPS and finite
size scaling phase boundaries are supported by an approximate analytical result using
a truncated Jordan-Wigner transformation approach in combination with the Kitaev model.
We introduced the dynamics of the \LRQI{} model through local
and global observables during a quench from the paramagnetic phase into
the ferromagnetic and antiferromagnetic phases crossing the critical point. We
related those measurements to the actual quantum critical point through
the maximum of the gradient of average global magnetization in the $x$-direction
and analyzed how this is affected by the total time of the quench. Finally, we
evaluated the scaling of defects in the Kibble-Zurek scenario leading to
the result of a slowly increasing (constant)  defect density at the end for increasing quench velocity
in comparison to the nearest-neighbor limit of the ferromagnetic
(antiferromagnetic) \LRQI{} model. This is reflected in the exponent of the
Kibble-Zurek scaling, e.g. in the ferromagnetic case we observe exponents
in the range from $\mu = 0.45$ to $\mu  = 0.6$ for power law exponents
$\alpha > 1.5$, where $\mu = 0.5$ is the limit of the nearest-neighbor
\QI{} model in the thermodynamic limit.

Having concluded our study, we point out open problems that can be addressed in future
research. As indicated by the list of the recent studies concerning the dynamics
of the \LRQI{} model \cite{Schachenmayer2013,Richerme2014,Hauke2013,Santos2016},
static physics has been well explored \cite{Dutta2001,Koffel2012,Vodola2015}, but
near-equilibrium dynamics such as the Kibble-Zurek hypothesis
remain largely unknown, let alone far-from-equilibrium dynamics. One aspect which could
be explored in statics and dynamics is the extension to two or higher
dimensional systems which we have excluded due to the limitation of MPS
algorithms to one spatial dimension for studies of this kind. Two-dimensional
setups allow more options for quantum computation to apply gates
\cite{Boothby2016}. In the case of the antiferromagnetic case triangular
lattices can lead to the similar concurrence of the staggered order than for
long-range interactions in the one-dimensional case.
While two-dimensional systems might be accessible through simulations
with tensor network methods such as Projected
Entangled Pair States (PEPS) \cite{Verstraete2008}, this route seems to be
more accessible for experiments for any of the models mentioned before. Discrete
Truncated Wigner methods are another tool to approach two-dimensional systems
with numerical simulations \cite{Schachenmayer2015}.
A further path starting from here is the investigation of dynamical phase
diagrams. Where recent studies treat e.g. binary pulses in regards to
interaction and transverse field of the \QI{} model \cite{vonKeyserlingk2016},
further possibilities considering time-dependent periodic long-range
interactions arise. One study obtaining a dynamical phase diagram for the
long-range quantum Ising chain in the infinite limit using iMPS is
\cite{Halimeh2016}. So called
completely interacting models at $\alpha = 0$ form another topic for research.

We have shown that the \LRQI{} model is a good starting
point for any kind of long-range study. Other models may show new behaviors when
including long-range interactions. We think here of the generalization of the
\QI{} in terms of the $XY$-model with a realization in Rydberg systems
\cite{Barredo2015} or the $XYZ$-model as realized in polyatomic molecules
\cite{Wall2015Top}. Furthermore, this includes Hubbard models such as the
Bose-Hubbard model, Fermi-Hubbard model as they appear e.g. for molecules. For
macroscopic quantum tunneling in the Bose-Hubbard model \cite{Diego2016}
results might change if the distance of the long-range interaction is larger
than the width of the barrier. Other problems suitable for extension to
long-range interactions include the quantum rotor model and the bilinear
biquadratic spin-one Hamiltonian \cite{Manmana2011}.

\section*{Acknowledgments}

We acknowledge useful discussions with Jad~C.~Halimeh, Gavriil~Shchedrin,
and David~L.~Vargas.
This work has been supported by the National Science Foundation under grant
numbers PHY-120881 and PHY-1520915 and the AFOSR grant FA9550-14-1-0287.
The calculations were carried out using the high performance computing
resources provided by the Golden Energy Computing Organization at the
Colorado School of Mines.

\section*{References}


\appendix

\section{Numerical Methods for Long-Range Statics and Dynamics                  \label{sec:methods}}

In this section we describe the numerical algorithms used, which can be
skipped by readers not interested in the computational aspects of this work.
The motivation for using tensor network methods for the simulation
of our systems is their versatile applications for finite and infinite size
statics and finite systems dynamics. Since the invention of the finite size
version of MPS \cite{Vidal2003}, various tensor network methods
have been proposed. These types of tensor networks include for pure states
Tree Tensor Networks (TTNs) \cite{Shi2006}, Multiscale Entanglement
Renormalization Ansatz (MERA) \cite{Vidal2008}, and Projected Entangled Pair
States (PEPS) \cite{Verstraete2004b}.
All approaches are based on a parameterization of states in the Hilbert space in
order to capture many-body physics which is well beyond the reach of exact
diagonalization. For an overview we refer to \cite{Schollwoeck2011}.
We divide our description of the specific methods used into three parts.
First, we discuss the convergence of the iMPS algorithm used to determine the phase
boundary in terms of the
critical external field $\hc$ as a function of the interaction decay $\alpha$.
This algorithm is based on variational methods to find the ground state of
a translationally invariant unit cell. Second, we treat the path to finite size
simulations via MPS returning the ground state of a system of size $L$.
Iterating over different system sizes $L$, we discuss how to obtain the critical
value of the external field $\hc$ in the limit $L \to \infty$ via finite size
scaling.
Third, we complete the section with a discussion of the dynamic
results obtained with the time-dependent variational principle \cite{Haegeman2016}
which is able to intrinsically capture long-range interactions and
is therefore preferable over a Suzuki-Trotter decomposition as used in many
implementations of MPS/TEBD libraries.

\subsection{Static Simulations: Infinite Systems and the Finite Size Scaling
  Approach                                                                      \label{sec:statics}}


We explain the numerical methods behind the simulation data shown in
Fig.~\ref{fig:PhaseBoundaries}. The iMPS algorithm
is designed to find the ground state of an infinite system
without boundary effects, returning the translationally invariant unit cell of
size $L$. New sites are inserted in the middle of the existing system and
variationally optimized in order to converge to the ground state.
The previous iteration is included as the \emph{environment},
where environment is understood in the sense that it exchanges quantum numbers
with the system, and not in an system-bath
open quantum system context. In the first step of the algorithm, the unit
cell is optimized without any environment; in the second step the solution
of the first step is considered as the environment. From that point on the
environment grows subsequently with each step. The algorithm terminates
if the newly introduced sites fulfill the orthogonality fidelity condition:
\begin{eqnarray}                                                                \label{eq:fortho}
  \mathcal{F} \left(\rho_{n-1}, \rho_{n}^{R} \right)
 &=& \mathrm{Tr} \sqrt{\sqrt{\rho_{n}^{R}} \rho_{n-1} \sqrt{\rho_{n}^{R}}} \, ,
\end{eqnarray}
where $n$ indicates the iteration step and $R$ is the right part of the
system excluding the site introduces in the most recent step. Therefore,
the density matrices $\rho_{n-1}$ and $\rho_{n}^{R}$ represent the same
part of the system and the latter one has one more update included.
When the overlap is sufficiently big, the update did not change the result
any more.
We refer to \cite{McCulloch2008,WallNJP2012} for a detailed description.
The phase boundary for the infinite MPS simulation is obtained for each
value of $\alpha$ evaluating the maximum of the bond entropy iterating
over the external field $h$. The bond entropy is defined by the von Neumann
entropy introduced in Eq.~\eqref{eq:vonNeumann}. We obtain the singular
values squared $\lambda$ when splitting the unit cell of the infinite MPS
into the bipartition.
The bond entropy is one of a variety of possible entanglement measures
\cite{NielsenChuang,Plenio2007}, and it is based on the Schmidt decomposition
separating a quantum system into bipartitions. The
entanglement can be measured via the number of singular values, the
Schmidt number, or the entropy of the singular values according to Eq.~\eqref{eq:vonNeumann}. For example, a Bell
state has Schmidt number $2$ and von Neumann entropy $-2 \cdot (\frac{1}{2} \log(\frac{1}{2}))$,
while an unentangled product state has Schmidt number $1$ with von Neumann
entropy $0$. The maximum of the bond entropy is found
using a grid of $21$ points ranging from $h_{\min}^{[1]} = 0.9$
($h_{\min}^{[1]} = 0.0$) to $h_{\max}^{[1]} = 3.9$ ($h_{\max}^{[1]} = 2.0$) in the
ferromagnetic (antiferromagnetic) long-range Ising model. We run this
evaluation for $51$ different values of $\alpha$ ranging from $\alpha = 0$ to
$\alpha = 4$. The critical value
$\hc^{[1]}$ is the grid point with the maximal bond entropy. Based on this
result, we refine the search on the next grid defined as
\begin{eqnarray}
  \hc^{\lbrack i \rbrack} \Longrightarrow
  h_{\mathrm{low}}^{\lbrack i+1 \rbrack} =
     \hc^{\lbrack i \rbrack} - dh^{\lbrack i \rbrack} \, ,
  h_{\mathrm{up}}^{\lbrack i+1 \rbrack} =
     \hc^{\lbrack i \rbrack} + dh^{\lbrack i \rbrack} \, ,
\end{eqnarray}
where $dh^{[i-1]}$ is the step size on the grid. The results presented in
Sec.~\ref{sec:phaseboundary} and in the following convergence
studies are for $\hc^{[2]}$, that is the external field at the quantum critical
point between the paramagnetic and ferromagnetic, or antiferromagnetic phase.
The superscript denotes the iteration refining the grid for the search and has
no physical meaning past an indication for the precision of $\hc$.

\begin{figure}[htbp]
  \centering
  \includegraphics[width=0.99 \columnwidth]{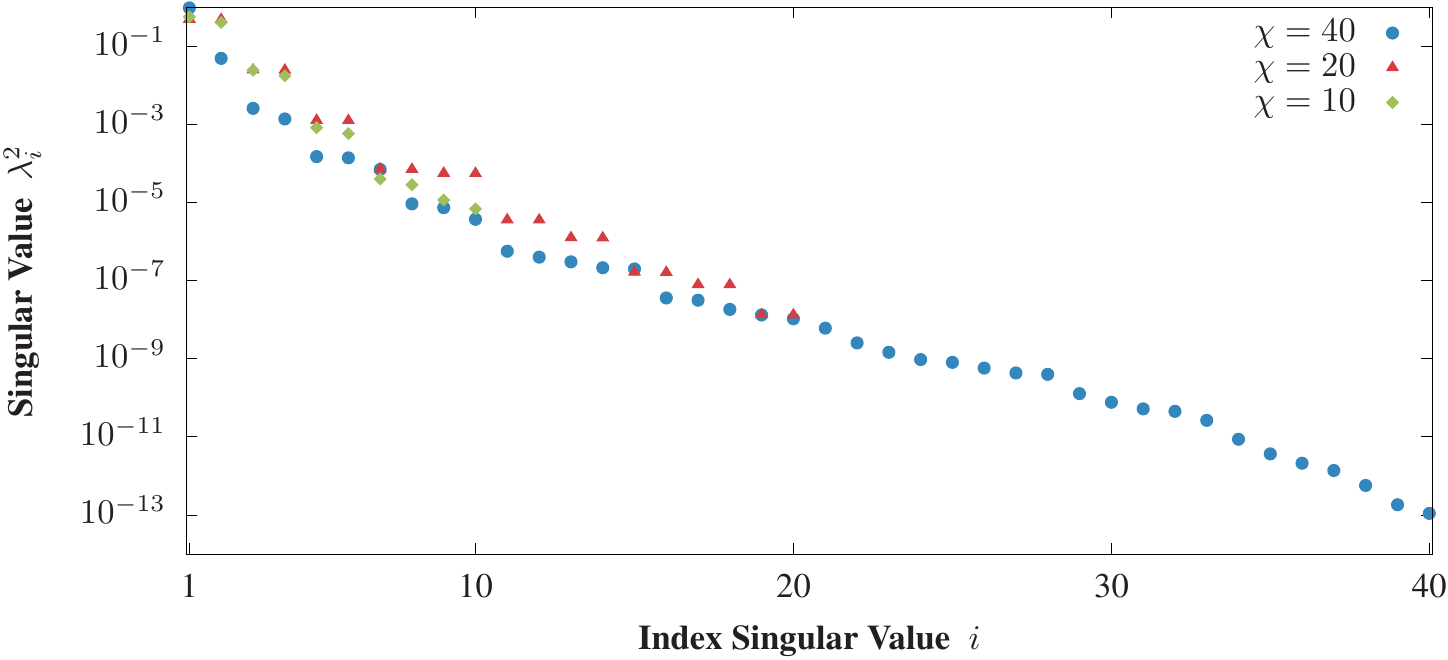}
  \caption{\emph{Degenerate Singular Values in iMPS.} The key for understanding
    the convergence of the bond entropy is $\Lambda$, the singular values
    squared, shown for different bond dimensions $\chi$.
    The difference in entropy is here related to the symmetry breaking appearing
    for $\chi=40$. In contrast the singular values for
    $\chi=10$ and $\chi=20$ appear in pairs. Considering the convergence with
    regards to $\chi$ by means of the bond entropy fails for this reason. This
    example represents a simulation for $\alpha = 4$ and $h = 1.14$ in the
    ferromagnetic model.
                                                                                \label{fig:iMPSZ2}}
\end{figure}

\begin{figure}[htbp]
  \centering
  \includegraphics[width=0.99 \columnwidth]{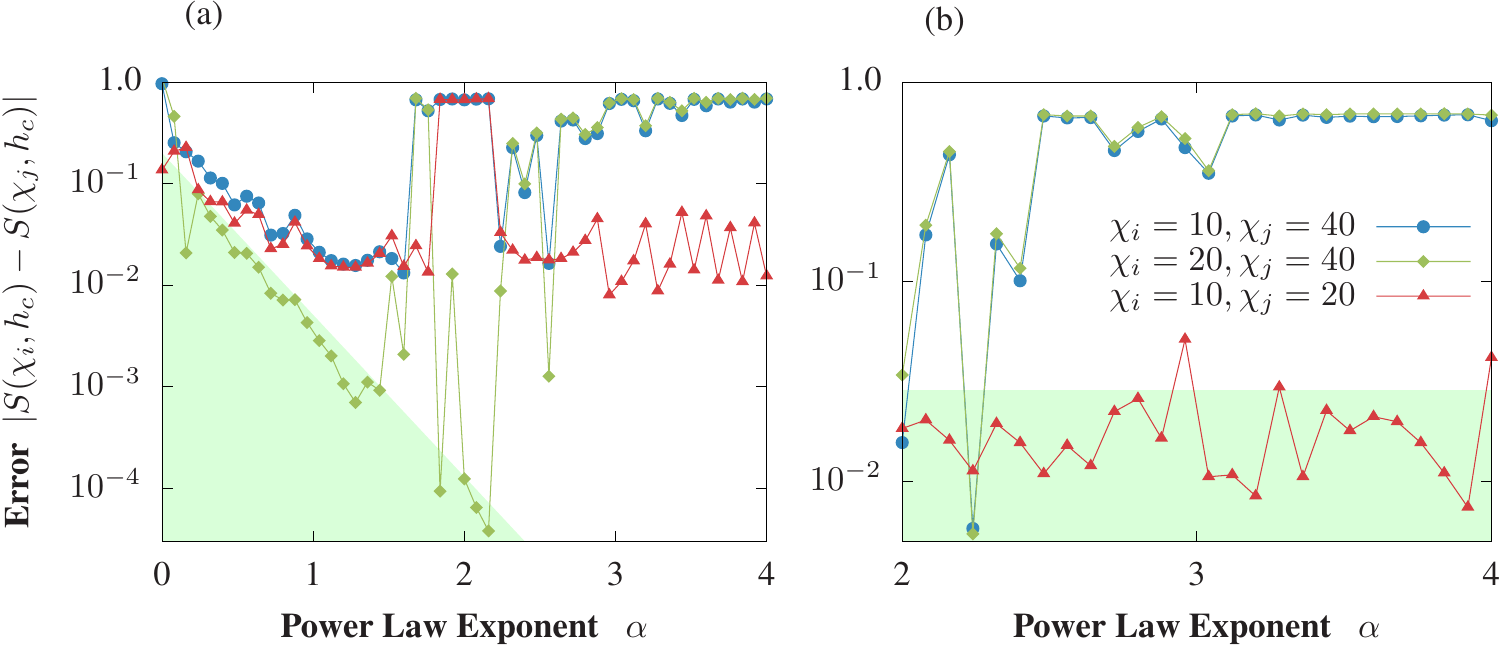}
  \caption{\emph{Convergence of iMPS.} (a) in the antiferromagnetic model on
    the left part of the plot we see the expected behavior that the error is
    increasing for longer-range interactions. This is indicated by the shaded
    green triangle, where values within the triangle have broken the symmetry
    or not for both bond dimensions $\chi = 20$ and $\chi = 40$. In contrast
    on the right part of the plot we have one simulation with conserved symmetry
    and one with broken symmetry for the comparison between $\chi = 20$ and
    $\chi = 40$.
    (b) In the range shown in the plot for the ferromagnetic case, the
    degeneracy is never kept up to $\chi = 40$ so that the error
    $| S(\chi=40) - S(\chi=20) |$ is not a reasonable basis on which to judge
    convergence. But the simulations comparing $\chi = 10$ and $\chi = 20$
    in the shaded green rectangle are a meaningful error with regards to the
    iMPS symmetry breaking issue.
    Legend applies to both (a) and (b).
                                                                                \label{fig:StaticIMPSConv}}
\end{figure}

We now discuss convergence of the iMPS
results. We take the default convergence parameters of the \OSMPS{} library \cite{openMPS}
with the modification of an iterative bond dimension of $\chi = 10, 20, 40$. In
Fig.~\ref{fig:StaticIMPSConv} (a) we show the difference in bond entropy
$|S(\chi=40, \hc) - S(\chi=20, \hc)|$, $|S(\chi=40, \hc) - S(\chi=10, \hc)|$,
and  $|S(\chi=20, \hc) - S(\chi=10, \hc)|$. These results call for a technical
discussion of this behavior due to the large difference for many
points with $\alpha \gtrsim 2$, here in the antiferromagnetic case. This behavior
is related to symmetry breaking. The degenerate ground state in the limit
$h \to 0$ is the GHZ state
\begin{eqnarray}                                                                \label{eq:Z2GHZ}
  \ket{\psi_0} = \ket{\psi_\mathrm{GHZ}}
 &=& \frac{\ket{\uparrow \cdots \uparrow}
     \pm \ket{\downarrow \cdots \downarrow}}{\sqrt{2}} \, ,
\end{eqnarray}
but without enforcing $\mathbb{Z}_{2}$ any superposition of
$\ket{\uparrow \cdots \uparrow}$ and $\ket{\downarrow \cdots \downarrow}$
minimizes the energy of the system. Therefore, the bond entropy can be
between the bond entropy $S = 0$ of a product state and
$S = \log(2) \approx 0.69$ of the GHZ ground state.
We discuss the origins of this problem related
to the numerical algorithm of iMPS. Tracking the singular values, we find
either a distribution of pairs in the states with high entropy corresponding
to the two different signs as pointed out in the limit $h \to 0$ for the GHZ
state in Eq.~\eqref{eq:Z2GHZ}. The low entropy states have decaying
non-degenerate singular values. The iMPS algorithm selects one of the
degenerate states at non-predictable points due to the randomized initial
state at the very beginning of the algorithm, as well as the entanglement
truncation step, which can randomly pick a particular parity sector for the
environment. This is related to the dominant eigenvector of the transfer
matrix (see \cite{McCulloch2008}). Note that, once one of the broken symmetry
states $| \uparrow \cdots \uparrow \rangle$ or $| \downarrow \cdots \downarrow
\rangle$ becomes more dominant, this dominance is "stored" for all subsequent
iterations in the environment in the iMPS algorithm. Hence, symmetry breaking
by random random numerical noise cannot be removed by running more iterations.
The higher bond dimension and
the number of iterations in the algorithm lead to smaller differences in the
pairs of degenerate singular values until one of the dominant eigenvectors
of the transfer matrix is kept and the other truncated. For $\chi = 10$, the
difference in the leading singular values is $\approx 0.15$ decreasing to
$\approx 0.05$ for $\chi = 20$. Therefore, the singular values double in
magnitude since an equal counterpart was neglected. In summary we can state the following:
(1) For $\alpha \lesssim 1.3$ symmetry breaking is not resolved. The
result converges, as expected, better for faster decaying long-range
interaction up to $10^{-3}$ between the results for $\chi=20$ and $\chi=40$.
(2) For $a \gtrsim 2$ we truncate one of the symmetry broken states for
$\chi = 40$ meaning that the difference in entropy is large for any comparison
to $\chi = 40$. The results with bond dimension $20$ have an error
between $10^{-1}$ and $10^{-2}$ with regard to $\chi = 10$. The singular values
for $\chi=40$ are close enough to be resolved as degeneracy leading to
the conclusion that the state is also well converged.
(3) The region in between the algorithm alternates between resolving the
symmetry broken ground state or not, leading to convergence up to $10^{-4}$.
The same trend can be stated in the ferromagnetic case shown in
Fig.~\ref{fig:StaticIMPSConv} (b). A corresponding example for the
degeneracy evolving in the singular values can be seen in
Fig.~\ref{fig:iMPSZ2}. We emphasize that this analysis is relevant if searching
for the maximal entropy on a very coarse grid or when trying to show convergence
across bond dimensions. For a single bond dimension we may use the final value
of the orthogonality fidelity described in Eq.~\eqref{eq:fortho}. The convergence
in terms of the orthogonality fidelity is discussed in \ref{sec:convimps}.

\begin{figure}[htbp]
  \centering
  \includegraphics[width=0.99 \columnwidth]{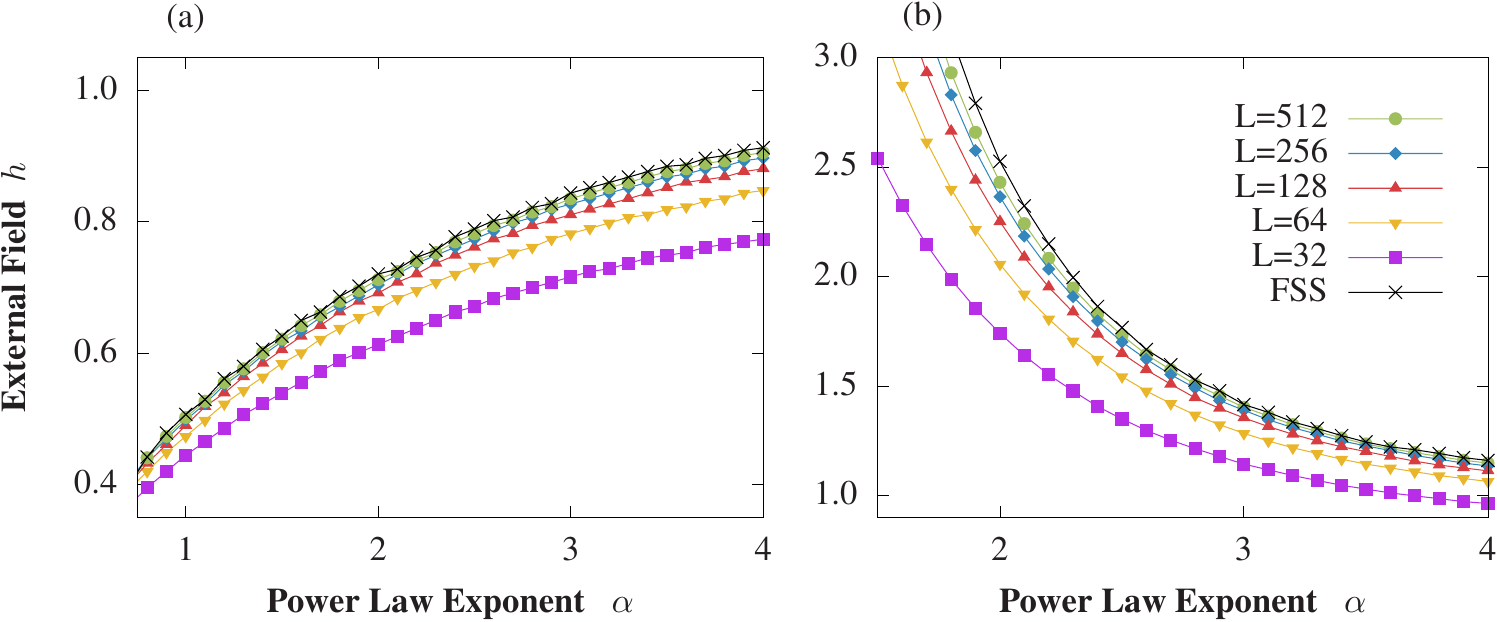}
  \caption{\emph{Finite Size Scaling for the \CLRQI{} Model.}
    Finite-size effects lead to a decrease of the
    critical value of the external field $\hc$ for smaller system sizes.
    Furthermore we achieve the thermodynamic limit for infinite system size
    via finite size scaling (FSS). (a) antiferromagnetic case; (b)
    ferromagnetic case. Legend applies to (a) and (b) both. We point
    out that for the limit $\alpha \to 0$ and $h \to 0$ the ground state for
    the ferromagnetic case is always the GHZ state, while the antiferromagnetic
    case has an exponential degeneracy. The latter might contribute to
    convergence problems when $\alpha$ is approaching zero. The limits
    $\alpha \to \infty$ is equal to the nearest-neighbor case and all limits
    $h \to \infty$ have a ground state aligned with the external field.
                                                                                \label{fig:StaticCritFss}}
\end{figure}

The finite size simulations have a two-fold purpose. On the one hand,
finite size scaling provides an alternate route to obtain the thermodynamic limit
of the system and is therefore a check against the iMPS results. While iMPS
simulations are only limited by the bond dimension used, finite size
simulations enforce the $\mathbb{Z}_{2}$ symmetry and avoid for this reason the
symmetry-breaking effects discussed in the previous section.
On the other hand, we see the finite size
effects affect the system, which is important as the dynamic simulations are
for finite systems.
For the finite size scaling we use the same grid method as for iMPS before
with the initial grid boundaries $h_{\min}^{\lbrack 1 \rbrack} = 0.8$
($h_{\min}^{\lbrack 1 \rbrack} = 0.0$) and $h_{\max}^{\lbrack 1 \rbrack} = 9.8$
($h_{\max}^{\lbrack 1 \rbrack} = 2.0$) with nine (seven) grid points for each
$\alpha$ and $L$ in the ferromagnetic (antiferromagnetic) case. We refine the
grid four times after the initial run ending with the bond dimensions
$\chi = (256, 512)$ for the iteration over the \OSMPS{} convergence parameters. The final distance
between points is for both cases $dh^{[5]} \approx 0.004$. For the refinement we
find $\hc^{\lbrack i \rbrack}(\alpha, L)$ with the bond entropy at the
middle of the system and construct the next, refined, grid $i + 1$ in the interval
$\hc^{\lbrack i \rbrack}(\alpha, L) \pm dh^{[i]}$. The maximum number of sweeps
are $(4, 6)$ for the corresponding iterations in bond dimension
$\chi = (256, 512)$. The other convergence parameters, which stay constant for
each set of convergence parameters iterated over, are a variance tolerance of
$10^{-10}$ and a local tolerance of $10^{-10}$. For power law exponents of
$\alpha > 1.5$, we actually reach a variance superior to $10^{-7}$ where
simulations converge better away from the critical value for the ferromagnetic
Hamiltonian. In the antiferromagnetic case, longer-range interactions show worse
convergence independent of the critical value of the external field. We discuss
the details in \ref{sec:convmps}.
The critical values are evaluated for system sizes $32$, $64$, $128$, $256$,
and $512$.
By using finite size scaling, we obtain the critical value $\hc$ of
an infinite system. From Cardy \cite{Cardy2003} we know that for a finite
system the distance of the critical point $\hc(L)$ to the critical point in
the thermodynamic limit $\hc$ can be described by
\begin{eqnarray}
  \left | \hc(L) - \hc \right | \propto L^{-\frac{1}{\nu}} \, . 
\end{eqnarray}
Knowing that the values of $\hc$ are increasing in the ferromagnetic and
antiferromagnetic case according to Fig.~\ref{fig:StaticCritFss},
we can use
\begin{eqnarray}
  \hc(L) = \hc - c L^{-\frac{1}{\nu}} \, ,
\end{eqnarray}
and fit the unknown parameters $\hc$, $\nu$, and $c$ via \cite{scipy}. Due to the
logarithmic choice of the system size, small system sizes are overrepresented
in the fit. We account for this by assigning an uncertainty proportional
to $1 / L$ at each data point. The results for different system sizes and
the data from the finite size scaling can be seen in
Fig.~\ref{fig:StaticCritFss}. In the antiferromagnetic case (a), the points
show the expected behavior with smaller spacing between bigger system sizes;
thus the simulations are converging to the thermodynamic limit. When the
interactions are not decaying for $\alpha \to 0$, the finite size scaling breaks
down due to poorly-converging simulations. The black curve represents the
$L \to \infty$ limit achieved through the finite size scaling fit.
Figure~\ref{fig:StaticCritFss} (b) shows the same behavior for the
ferromagnetic case. We point out that the weights for each data point are
necessary in this case to avoid an intersection of the curve for $L \to \infty$
with the finite size solution $L=512$ for longer-range interactions.

\subsection{Dynamic Simulations with MPS                                        \label{sec:dynamics}}

In this section we analyze the scaling of the numerical simulation with regards
to the Kibble-Zurek hypothesis. The Kibble-Zurek mechanism describes the
scaling of the defect density when driving a quantum system through a quantum
phase transition and was originally formulated in terms of cosmology
\cite{Kibble1976}, and later brought to quantum systems such as helium
\cite{Zurek1985} or the nearest-neighbor \QI{} model \cite{ZurekPRL2005}. There,
the external field is changed in order to cross the quantum critical point at
$\hc = 1$ in the nearest-neighbor \QI{} model. In our analysis the quantum
critical point $\hc$ depends on the actual value of the power law exponent
$\alpha$ as presented in Sec.~\ref{sec:ham}.

In order to analyze the Kibble-Zurek hypothesis in the present of long-range
interactions, we first define the defect
density for the ferromagnetic model. The defect density was defined in
\cite{DziarmagaPRL2005} using nearest-neighbor correlations, and we divide
by $(L-1)$ instead of $L$ to account for the open boundary condition:
\begin{eqnarray}                                                                \label{eq:defectdensity}
  \rho &=& \frac{1}{2 (L - 1)} \sum_{i=1}^{L-1}
  \left(1 - \langle \sigma_{i}^{z} \sigma_{i+1}^{z} \rangle \right) \, .
\end{eqnarray}
We now shortly outline why this works in the ferromagnetic case for the
nearest-neighbor model. We recall that at the end of the quench the external
field $h=0$ and therefore the ground states $\ket{\psi_{0}^{\mathrm{F}}}$ are
defined through
\begin{eqnarray}
  \ket{\psi_{0}^{\mathrm{F}}} =
  \frac{1}{\sqrt{2}} \left( \ket{\uparrow \ldots \uparrow} \pm
                            \ket{\downarrow \ldots \downarrow} \right) \, .
\end{eqnarray}
This is equivalent to the GHZ state and we recall that this is the ground
state before $\mathbb{Z}_2$ symmetry breaking. The first excited state has $2 (L - 1)$
degeneracies and is described by a single
kink in the system, e.g. $\ket{\ldots \uparrow \uparrow \downarrow \downarrow
\ldots}$. The number of kinks is equal to the number of
excitations. If we consider first a state with a product state of only spins up
or down, one is easily convinced that Eq.~\eqref{eq:defectdensity} describes
the defect density. For every domain wall the nearest-neighbor correlation
inside the sum contributes, while for spin inside the domain wall the term is
cancelled. Moreover, it is a
reasonable measurement even when the spins are changing slowly. Let us verify
this point with an example of three spins changing from up to down, $\ket{\phi} =
\ket{\uparrow} \ket{\uparrow}_{x} \ket{\downarrow}$. The subscript reflects
the basis if it differs from $z$. The measurement of the correlation yields
zero in both cases:
\begin{eqnarray}
  \bra{\uparrow} \bra{\uparrow}_{x} \sigma_{1}^{z} \sigma_{2}^{z}
     \ket{\uparrow} \ket{\uparrow}_{x} = 0 \, \qquad
  \bra{\uparrow}_{x} \bra{\uparrow} \sigma_{2}^{z} \sigma_{3}^{z}
     \ket{\uparrow}_{x} \ket{\uparrow} = 0                                      \nonumber \\
  \Longrightarrow \rho = \frac{1}{2L} \left( (1 - 0) + (1 - 0) \right)
    = \frac{1}{L} \times 1 \, .
\end{eqnarray}
The measurement in Eq.~\eqref{eq:defectdensity} yields one defect, as expected,
with the spins flipping once over the chain. We emphasize that this cannot be
used in any region with $h \neq 0$, except as an approximation for $h < \hc$.
Let us turn to the defects of the ferromagnetic \LRQI{} model. The defects
are local kinks, but the states with one kink are not all degenerate. In fact,
the energy gap to the ground state is smaller for kinks closer to the
boundaries. This is a manifestation of the boundary effects. Counting all
kinks equally, we make an error growing in the limit $\alpha \to 0$.
The defect density of the antiferromagnetic case in the nearest-neighbor model
is defined as
\begin{eqnarray}                                                                \label{eq:defectdensityaf}
  \rho_{\mathrm{AF}} &=& \frac{1}{2 (L-1)} \sum_{i=1}^{L-1}
  \left(1 + \langle \sigma_{i}^{z} \sigma_{i+1}^{z} \rangle \right) \, .
\end{eqnarray}
We again point out that the single defects have a different excitation energy
depending on their position in the bulk.

\begin{figure}[htbp]
  \centering
  \includegraphics[width=0.65 \columnwidth]{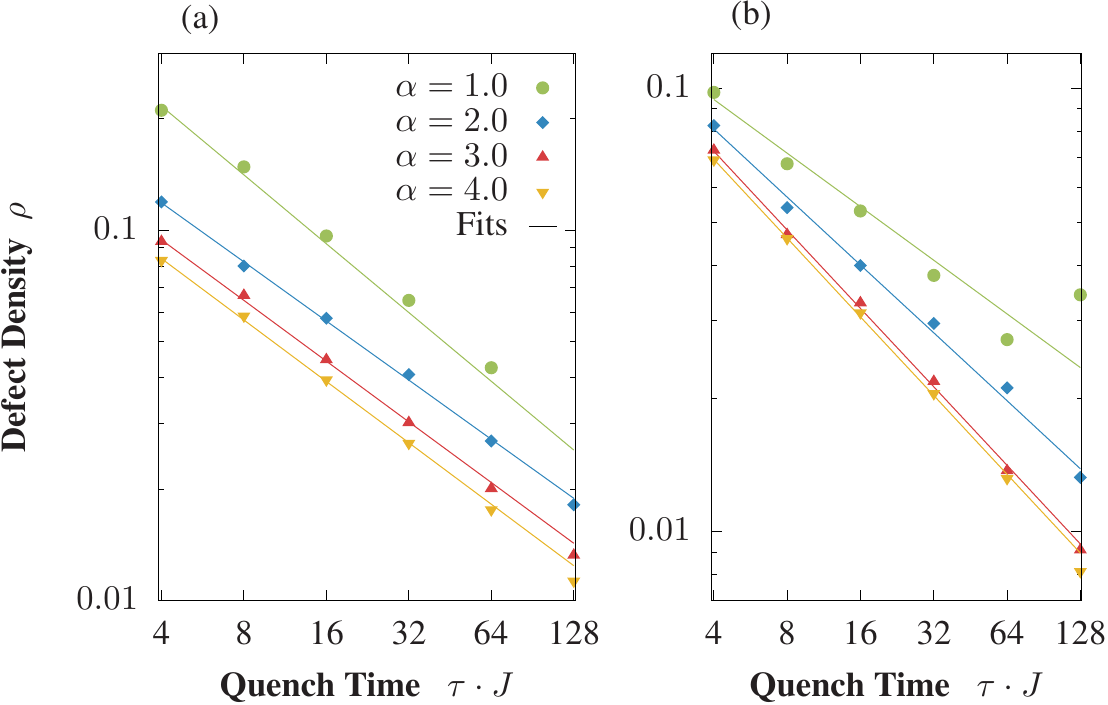}
  \caption{\emph{Fitting the Exponent for the Defect Density in the
    Kibble-Zurek Hypothesis.} (a) Examples for the defect density at the end of the
    quench for the antiferromagnetic \LRQI{} model plotted over the quench time
    $\tau$ for different power law exponents $\alpha$ shows the decrease in
    defect density for slower quenches. The power law in the scaling leads
    ideally to a line in the log-log plot which is here slightly affected by
    failing simulations. This is especially evident for small $\alpha$, e.g.
    in the antiferromagnetic case for $\alpha = 1.0$ and $\tau = 128$. (b)
    Corresponding data for the ferromagnetic model.
                                                                                \label{fig:KZMFit}}
\end{figure}

In order to retrieve the critical exponent scaling in the form of
Eq.~\eqref{eq:kzmscaling},
we simulate the quenches for $L=128$ for the set of $\tau$'s containing
$\{4, 8, 16, 32, 64, 128 \}$. For each $\alpha$ we fit the coefficient $c$ and
the critical exponent $\mu$ via \cite{scipy}. In order to estimate if any
non-converged simulations are included in the data, the standard
deviation $\sigma_{\mu}$ is calculated through the covariance matrix given
by the fitting function. In Fig.~\ref{fig:KZMFit} we show one example in the
ferromagnetic model for $\alpha = 1, 2, 3,$ and $4$. The error of the fit can
be seen as an indicator for the validity of the dynamic results. Running simulations with
different $\chi$ helps us to identify failing simulations versus singularities.
For weaker long-range
interactions with big $\alpha$ the fits reproduce the results while for
longer-range interaction the procedure fails ($\alpha=1$). The initial state
is based on the same convergence parameters as the statics discussed
in \ref{sec:statics}, except for the bond dimension which is lowered to
$\chi = 95$ or $100$. For the dynamics we keep the corresponding maximal bond
dimension and use the default TDVP convergence parameters in \OSMPS{}
\cite{openMPS} for all other parameters.

\section{Convergence of Finite-Size Statics                                     \label{sec:convmps}}

In the appendix we briefly address the convergence of the finite size statics.
We already discussed the convergence of iMPS in \ref{sec:methods} and
therefore it is known that the convergence can be studied in different
fashions. In \OSMPS{}, the variance of $H$ is considered for statics of finite
size systems. The user chooses a value, we take $10^{-10}$, and \OSMPS{}
iterates through the different bond dimension $\chi$. We consider $\chi = 256$
and $\chi = 512$ of the last grid. This fifth grid has a discretization of
$dh \approx 0.0044$ (ferro) and $dh \approx 0.0042$ (antiferro) around the
critical point of the fourth grid. The output contains the flag if the
simulation converged according to the value for each $\chi$; in addition the
actual value of the variance is saved as well.

\begin{figure}[htbp]
 \begin{center}
   \vspace{0.3cm}
   \begin{minipage}{0.49\linewidth}
     \begin{center}
     \begin{overpic}[width=0.98 \columnwidth,unit=1mm]{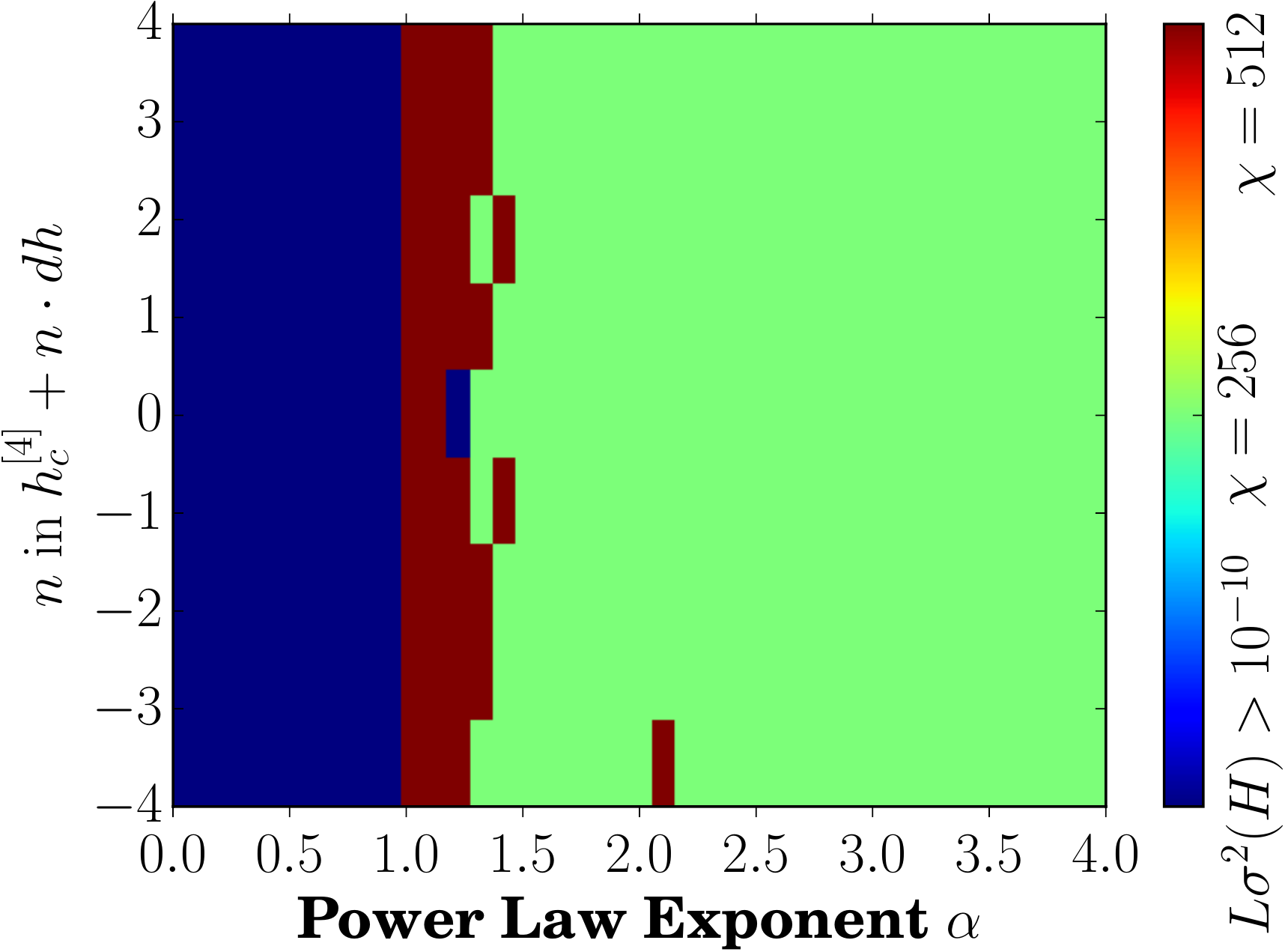}
       \put(12,77){(a)}
     \end{overpic}
     \end{center}
   \end{minipage}\hfill
   \begin{minipage}{0.49\linewidth}
     \begin{overpic}[width=1.0 \columnwidth,unit=1mm]{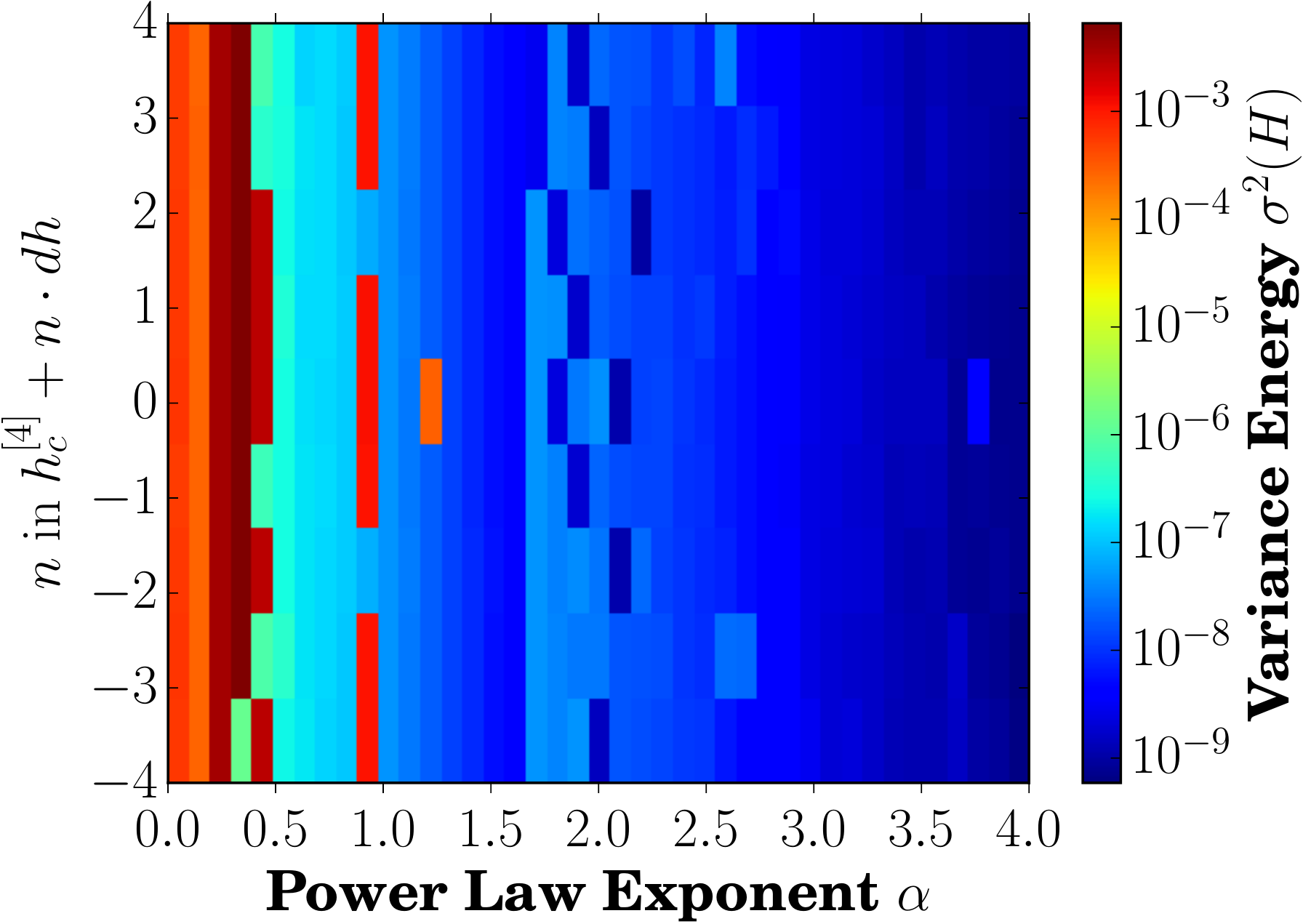}
       \put(12,74){(b)}
     \end{overpic}
   \end{minipage}\vspace{0.2cm}
  \caption{\emph{Convergence of Finite Size MPS Ground States:} built-in
    convergence check of \OSMPS{} using the variance of $H$ for the last
    grid and $L=512$ in the ferromagnetic case. (a) The critical region
    causes most problems to converge up to variance $L \cdot 10^{-10}$
    (green converged with $\chi=256$, red converged with $\chi=512$, blue not
    converged). (b) The actual variance of $H$ of the fifth
    grid for the results used in the main section is at least converged
    around $10^{-6}$.
                                                                                \label{fig:StatMPSConvF}}
  \end{center}
\end{figure}

We start to look if the simulation converged with a variance of $L \cdot 10^{-10}$.
This information is contained in the left part of Fig.~\ref{fig:StatMPSConvF}
for the ferromagnetic model. The coloring is as follows: green for convergence
at the first convergence parameter with $\chi = 256$, red for convergence the
second convergence parameter at $\chi = 512$, and blue if not converged at
either of the previous bond dimensions. In order to estimate the variance of the blue
region, we plot the actual value for the critical value $\hc$ as a function of
$\alpha$ and system size in the right part of Fig.~\ref{fig:StatMPSConvF}. For
the analysis used in the main part of the paper we achieve a variance less
than $10^{-7}$.

\begin{figure}[htbp]
 \begin{center}
   \vspace{0.3cm}
   \begin{minipage}{0.49\linewidth}
     \begin{center}
     \begin{overpic}[width=0.98 \columnwidth,unit=1mm]{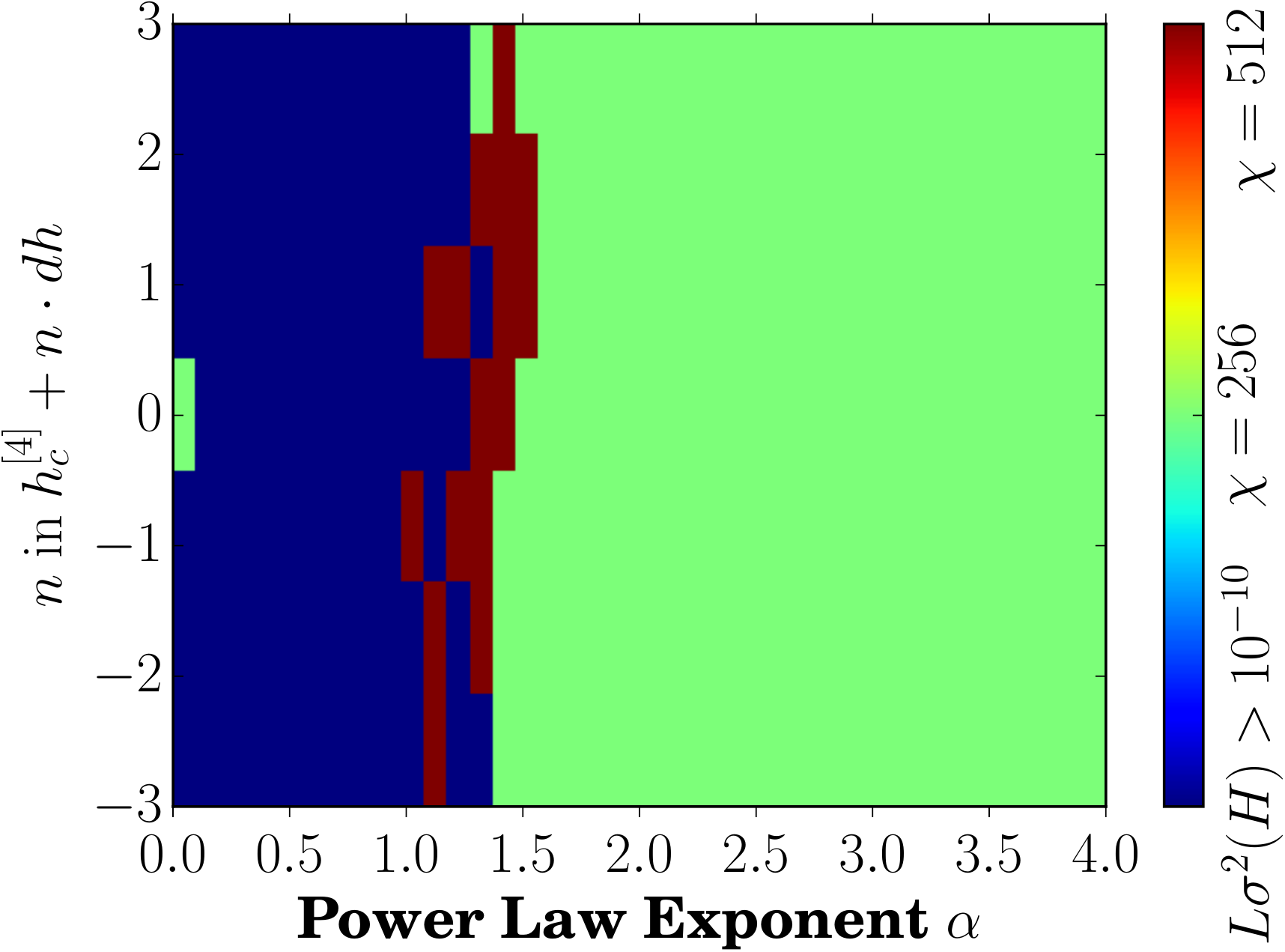}
       \put(12,77){(a)}
     \end{overpic}
     \end{center}
   \end{minipage}\hfill
   \begin{minipage}{0.49\linewidth}
     \begin{overpic}[width=1.0 \columnwidth,unit=1mm]{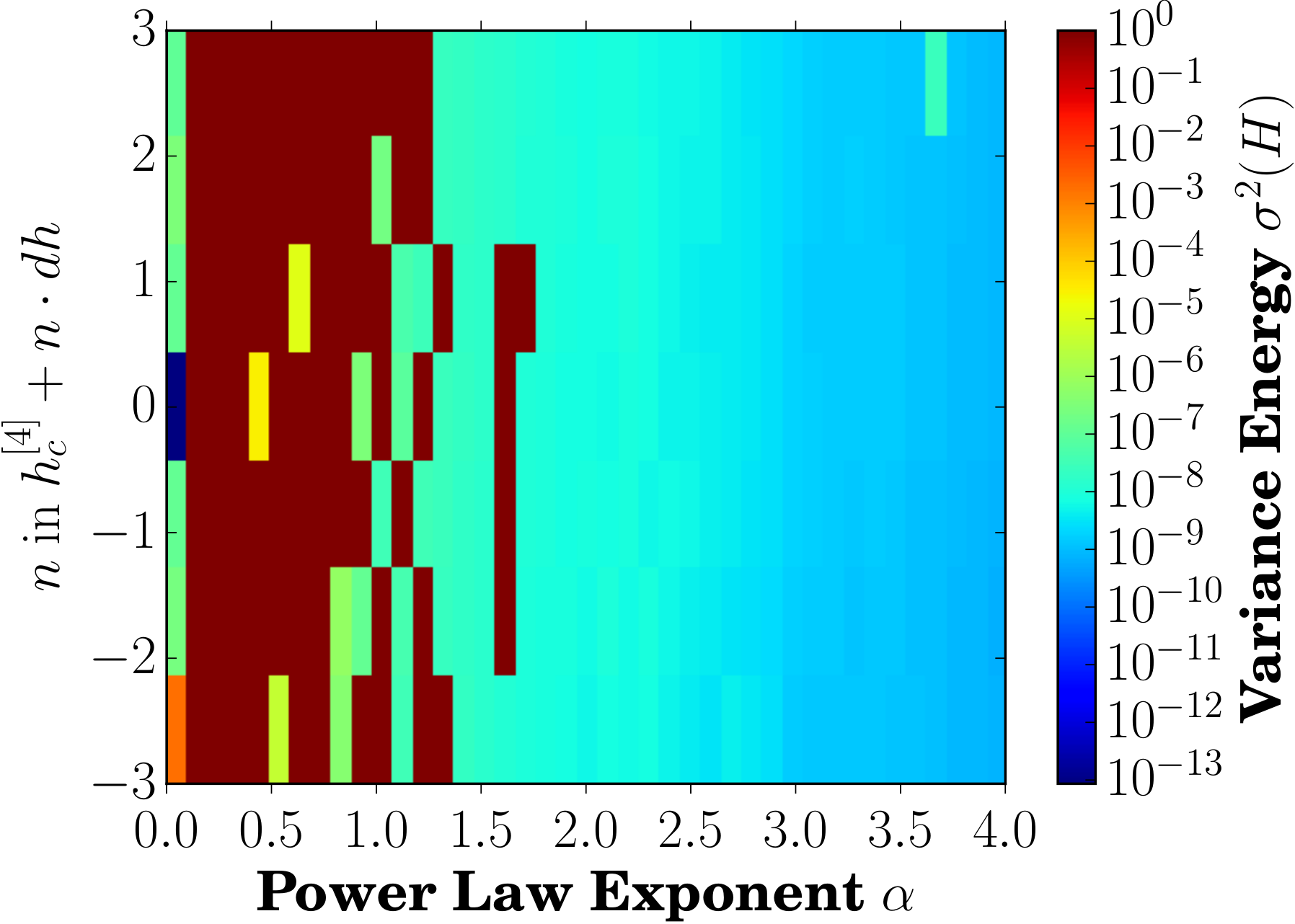}
       \put(12,74){(b)}
     \end{overpic}
   \end{minipage}\vspace{0.2cm}
  \caption{\emph{Convergence of Finite Size MPS Ground States:} built-in
    convergence check of \OSMPS{} using the variance of $H$ for the coarse
    grid and $L=512$ in the antiferromagnetic case. (a) The longer-range
    interaction region causes most problems to converge up to variance
    $L \cdot 10^{-10}$ (green converged with $\chi=256$, red converged
    with $\chi=512$, blue not converged). (b) The actual variance of $H$
    of the fifth grid is shown on the right. Some antiferromagnetic
    simulations did not finish in the allotted time.
                                                                                \label{fig:StatMPSConvAF}}
  \end{center}
\end{figure}

We follow the similar scheme to study the convergence in the antiferromagnetic
\LRQI{} Hamiltonian, presented in Fig.~\ref{fig:StatMPSConvAF}. The
convergence becomes in general more difficult in the longer-range interacting
region with $\alpha \to 0$, still reaching a variance of $10^{-6}$ or better
for the shorter-range region with $\alpha > 1.5$.


\section{Convergence of Infinite-Size Statics                                   \label{sec:convimps}}

\begin{figure}[htbp]
 \begin{center}
   \vspace{0.3cm}
   \begin{minipage}{0.49\linewidth}
     \begin{center}
     \begin{overpic}[width=1.0 \columnwidth,unit=1mm]{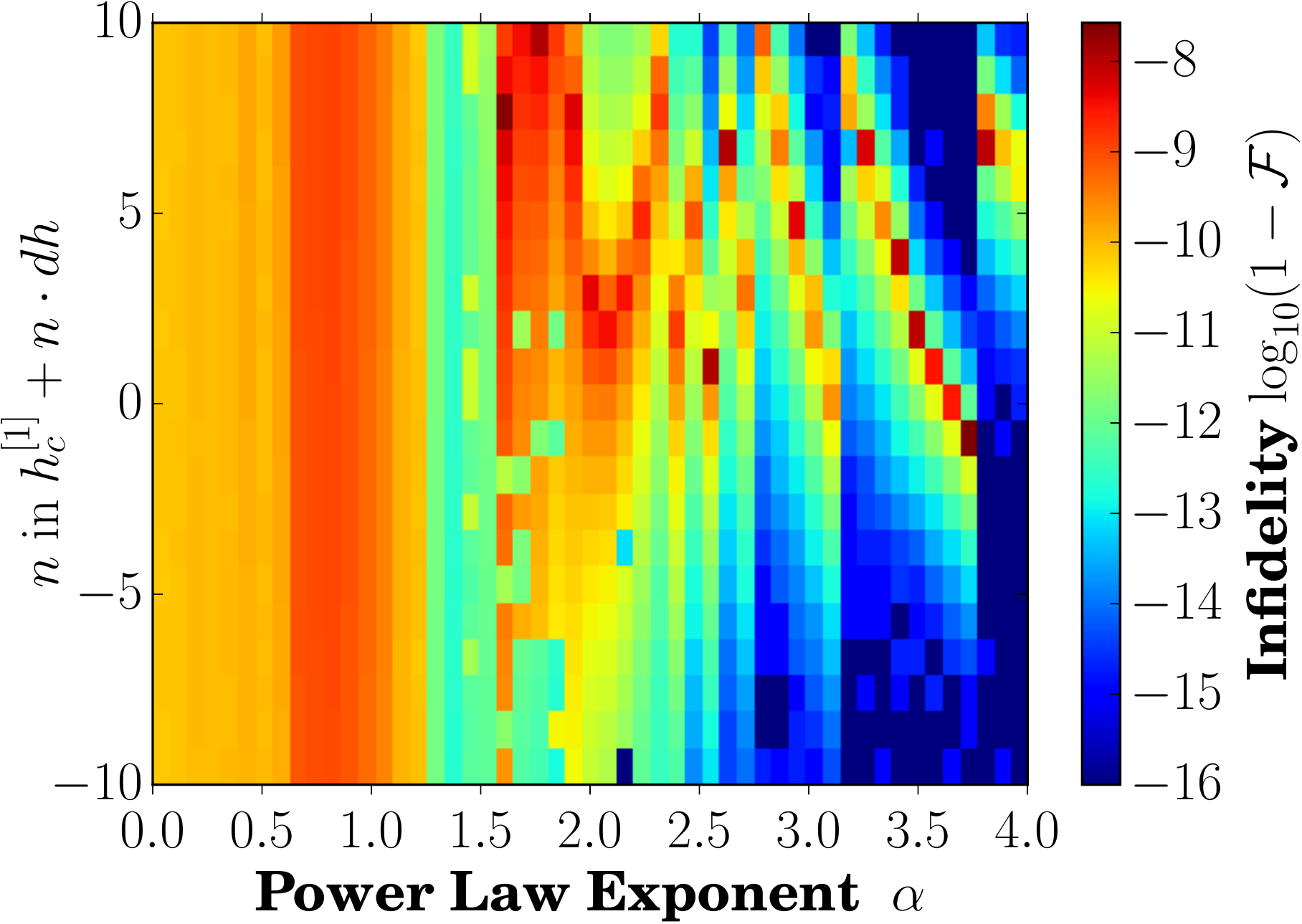}
       \put(12,74){(a)}
     \end{overpic}
     \end{center}
   \end{minipage}\hfill
   \begin{minipage}{0.49\linewidth}
     \begin{overpic}[width=1.0 \columnwidth,unit=1mm]{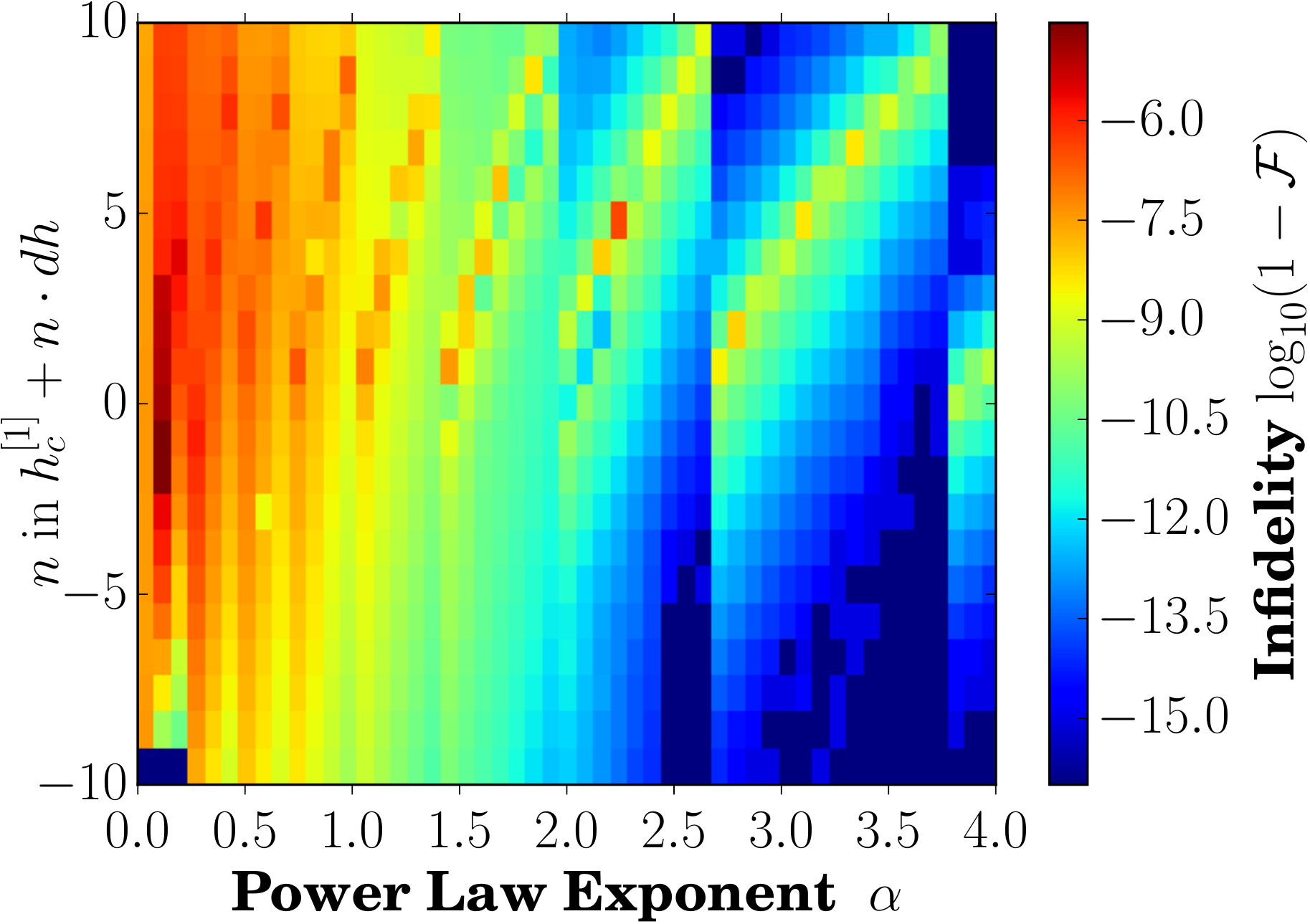}
       \put(12,74){(b)}
     \end{overpic}
   \end{minipage}\vspace{0.2cm}
  \caption{\emph{Convergence of the iMPS.} We consider the infidelity of the
    orthogonality fidelity plotted as logarithm of base 10. Both the
    ferromagnetic case (a) and the antiferromagnetic case (b) convergence
    for $\alpha > 1$ mostly to an infidelity of $10^{-8}$ or better.
    Non-converging points are typically around the critical point of the
    model moving downwards for the antiferromagnetic case and upwards for
    the ferromagnetic case. The vertical stripes originate in the fact
    that each simulation might have a different critical value from the
    first grid, where $\hc$ is identical within one strip.
                                                                                \label{fig:StatIMPSOrthoConv}}
  \end{center}
\end{figure}

We discussed at length that it is difficult to study convergence as a function
of the bond dimension in terms of the bond entropy in \ref{sec:statics}.
But we give in the following a brief overview of the convergence of the iMPS
in terms of the orthogonality fidelity defined in Eq.~\eqref{eq:fortho}. In
order to show convergence it is convenient to switch from the orthogonality
fidelity to the actual error, the infidelity
\begin{eqnarray}
  \mathcal{I} = 1 - \mathcal{F} \, .
\end{eqnarray}
Figure~\ref{fig:StatIMPSOrthoConv} shows the convergence for the
antiferromagnetic and the ferromagnetic model for the highest bond dimension
$\chi = 40$ of our convergence parameters and the second grid. In terms of
the orthogonality fidelity the simulations are well converged at an error
$\mathcal{I}$ of $10^{-8}$ or better for $\alpha > 1$. Convergence close to
the critical point is in general worse. For $\alpha < 1$ interaction become
too long-range to converge at the same level for these convergence
settings. We remind the reader that the search interval for the ferromagnetic
case shown in Fig.~\ref{fig:StatIMPSOrthoConv} (b) does not contain the
critical value for very small $\alpha$ with an upper bound of
$h_{\mathrm{upper}} = 3.9$.

\section{Finite Size Effects in the Kibble-Zurek Scaling                        \label{sec:fssdyn}}

Section~\ref{sec:dynamics} raised the question of how finite size effects
influence the value of the critical exponent $\mu$ in the dynamics. We consider
the quench times $\tau = 4, 8, 16, 32, 64, 128$ and different system sizes
$L = 32, 64, 128, 256$ to show these finite size effects in the nearest-neighbor
limit. This limit corresponds to $\alpha \to \infty$. Table~\ref{tab:fss_dyn}
presents the data. The error is solely from the fit and does not contain
truncation error or methological errors from the time evolution. The
thermodynamic limit is obtained via finite size scaling using the following
approach:
\begin{eqnarray}
  | \mu_{\infty} - \mu_L | \propto L^{-c_2} \Longrightarrow
  \mu(L) = \mu_{\infty} + c_1 L^{-c_2} \, .
\end{eqnarray}
The value $\mu_{\infty}$ is shown in the last column of
Table~\ref{tab:fss_dyn}. As in the finite size scalings previous discussed,
we weight the data points for large system sizes more since we have less data
points in the corresponding interval. We do not show the error here, because
of the limited number of data points with regards to the degrees of
freedom. In conclusion, the data supports the assumption that the finite
size effects are relevant and the thermodynamic limit in the nearest-neighbor
model approaches $\mu = 0.5$ in both the ferromagnetic and antiferromagnetic
case.

\begin{table}[t]
  \centering
  \begin{tabular}{@{} cccccc @{}}
    \toprule
    System size       &$L = 32$           &$L = 64$           &$L = 128$          &$L = 256$          &$L \to \infty$     \\
    \cmidrule(r){1-1} \cmidrule(rl){2-2}  \cmidrule(rl){3-3}  \cmidrule(rl){4-4}  \cmidrule(rl){5-5}  \cmidrule(rl){6-6}
    Anitferro         &0.810$\pm$0.060    &0.614$\pm$0.026    &0.532$\pm$0.016    &0.502$\pm$0.017    &$0.48$             \\
    Ferro             &0.810$\pm$0.060    &0.614$\pm$0.026    &0.532$\pm$0.016    &0.500$\pm$0.017    &$0.48$             \\
    \bottomrule
  \end{tabular}
  \caption{\emph{Finite Size Effects in the Kibble-Zurek Scaling for the
    Quantum Ising Model.} The finite size effects lead to a growing
    critical exponent for smaller system sizes. In the thermodynamic
    limit the critical exponent of $0.5$ is reached.
                                                                                \label{tab:fss_dyn}}
\end{table}

\end{document}